\documentclass[12pt]{article}
\usepackage{amsmath}
\usepackage{graphicx}
\usepackage{enumerate}
\usepackage{natbib}
\usepackage{url} 

\usepackage[section]{placeins}

\usepackage{amssymb,physics}
\usepackage{mathrsfs,bm,bbm}
\usepackage[svgnames]{xcolor}
\usepackage{tikz}
\usepackage{enumitem}
\usepackage{multirow,multicol}
\usepackage[ruled]{algorithm2e}

\usepackage{amsthm}
\newtheorem{theorem}{Theorem}
\newtheorem{lemma}{Lemma}
\newtheorem{corollary}{Corollary}
\theoremstyle{remark}
\newtheorem{remark}{Remark}

\theoremstyle{definition}

\usepackage[bookmarksnumbered=true, colorlinks,
	citecolor=ForestGreen,       
	linkcolor=Blue,		         
	urlcolor=Magenta,	     	 
]{hyperref}

\graphicspath{{./art/}}

\renewcommand{\Pr}{\mathbb{P}}
\newcommand{\E}{\mathbb{E}}
\newcommand{\Var}{\mathrm{Var}}
\newcommand{\Cov}{\mathrm{Cov}}

\newcommand{\1}{\mathbbm{1}}

\newcommand{\R}{\mathbb{R}}
\newcommand{\e}{\mathrm{e}}
\newcommand{\iid}{\textsc{iid}}

\begin{document}

\def\spacingset#1{\renewcommand{\baselinestretch}%
{#1}\small\normalsize} \spacingset{1}

\newcommand{\blind}{0}

\if0\blind
{
  \title{\bf A General Test for Independent and Identically Distributed Hypothesis}
  \author{Tongyu Li\\
    Department of Probability \& Statistics, School of Mathematical Sciences,\\ Center for Statistical Science, Peking University\\
    Jonas Mueller\\
    Cleanlab\\
    Fang Yao\thanks{
    Fang Yao is the corresponding author: \texttt{fyao@math.pku.edu.cn}. This research is partially supported by the National Natural Science Foundation of China (No. 12292981, 12288101), the National Key R\&D Program of China (No. 2022YFA1003801), the Newcorner Stone Science foundation through the Xplorer Prize, the LMAM and the Fundamental Research Funds for the Central Universities, Peking University (LMEQF).}\\
    Department of Probability \& Statistics, School of Mathematical Sciences,\\ Center for Statistical Science, Peking University}
    \date{}
  \maketitle
} \fi

\if1\blind
{
  \bigskip
  \bigskip
  \bigskip
  \begin{center}
    {\LARGE\bf A General Test for Independent and Identically Distributed Hypothesis}
\end{center}
  \medskip
} \fi

\bigskip
\begin{abstract}
We propose a simple and intuitive test for arguably the most prevailing hypothesis in statistics that data are independent and identically distributed (\iid), based on a newly introduced off-diagonal sequential U-process. 
This {\iid} test is fully nonparametric and applicable to random objects in general spaces, while requiring no specific alternatives such as structural breaks or serial dependence, which allows for detecting general types of violations of the {\iid} assumption. 
An easy-to-implement jackknife multiplier bootstrap is tailored to produce critical values of the test. 
Under mild conditions, we establish Gaussian approximation for the proposed U-processes, and derive non-asymptotic coupling and Kolmogorov distance bounds for its maximum and the bootstrapped version, providing rigorous theoretical guarantees. 
Simulations and real data applications are conducted to demonstrate the usefulness and versatility compared with existing methods. 
\end{abstract}

\noindent%
{\it Keywords:} 
Gaussian approximation; 
{\iid} test; 
multiplier bootstrap; 
off-diagonal sequential U-process
\vfill

\newpage
\spacingset{1.9} 

\section{Introduction}
\label{sec:intro}
The assumption that the data under study are independent and identically distributed (\iid) lays the foundation for many statistical learning procedures \citep{ganssler1979empirical,hsieh2020non,cao2022beyond}. 
Understanding and confirming at least partially the {\iid} property of data are essential for the validity and reliability of various statistical methods, such as maximum likelihood estimation and likelihood ratio tests \citep{rice2007mathematical,casella2024statistical}. 
In regression analysis and predictive modeling \citep{montgomery2021introduction,freedman2009statistical,clarke2018predictive}, complying to the {\iid} assumption can lead to appropriate inference and accurate prediction. 
A great deal of efforts in sampling \citep{fuller2009sampling} and experimental design \citep{wu2021experiments} have been devoted to acquiring {\iid} data, which sets stage for valid statistical analysis. 
Moreover, the {\iid} regime goes beyond statistics. For instance, in industrial settings, ensuring the consistency and stability of manufacturing processes is critical for maintaining product quality \citep{box2015time}. 

Due to its paramount role in modeling uncertainty, a vast literature on time series examines the {\iid} hypothesis. Often with special purposes, existing tests are designed against specific alternatives, e.g., structural breaks and serial dependence. 
The change-point problem has received attention with respect to distributional structures like mean and covariance \citep{aue2024state,madrid2022change,preuss2015detection,yu2022robust}. 
Pioneered by \citet{box1970distribution,ljung1978measure}, white noise testing that checks autocorrelation is a long-standing active research area \citep{dalla2022robust,fokianos2018testing,jiang2024testing}. 
Besides, \citet{klaassen2001points} appealed to an ad hoc linear probability model to test the {\iid} hypothesis. 
These methods inevitably undermine generalizability and lack of sensitivity to unsuspected violations of the {\iid} property. 
Few of the {\iid} tests take general alternatives into consideration \citep{cho2011generalized,gehlot2025evaluating}, but are only suitable for scalar variables and cannot be directly used for random objects in general spaces, such as images \citep{lecun1998gradient,krizhevsky2009learning} and networks \citep{ginestet2017hypothesis,dubey2022modeling}, which have become increasingly prevalent as technology advances. 
In view of this, we would like to propose a new approach to {\iid} testing that has great generality with practical and easy implementation.

Let the sample $X_{1:n}=\{X_1,\dots,X_n\}$ be a collection of random elements valued in a measurable space $\mathcal{X}$, which allows for accommodating complex types of data. 
The null hypothesis of interest is \[ H_0 : X_1,\dots,X_n \text{ are \iid}. \]
Here the ambient probability model and the alternative hypothesis are not burdened with prior information. 
Consequently, the proposed {\iid} test is fully nonparametric and flexible in use, enabling conclusions without stringent restrictions on the population. 
Nonparametric statistical inference has gained prominence over the past decades \citep{siegel1957nonparametric,hollander2013nonparametric}, and remains prosperously thriving in the era of big data. 
As an illustration, recent developments about nonparametric testing range from two-sample distributional comparisons \citep{xue2020distribution,hu2024two,kim2020robust,deb2023multivariate} to mutual independence tests \citep{chen2018testing,shi2022universally,wang2024nonparametric,bucher2024testing,zhou2024rank}, which can also be seen as handling certain aspects of the {\iid} hypothesis. 
Nevertheless, the issue of {\iid} testing is far from well researched, despite its immense importance. 
Another line of relevant literature considered testing exchangeability in an online setting that focused on an infinite sequence of observations \citep{vovk2021testing,saha2024testing}.  
The nature of sequential testing therein allows for online change detection, where the task is to raise an alarm soon after the assumption of exchangeability becomes violated. 
By contrast, our null hypothesis $H_0$ is primarily concerned with the offline setting that is a more standard practice in statistical analysis. 
It is worth mentioning that exchangeability is a weaker interpretation of randomness and should be distinguished from the {\iid} property in the setting of finitely many observations \citep{diaconis1980finite,lefevre2017finite}. 

We tackle the challenge of testing the {\iid} assumption for random objects $X_{1:n}$ through an elaborately devised test statistic, which may shed new light on modern data analysis. 
The proposed framework bridges exploratory diagnostics and confirmatory testing by exploiting a pivotal insight: under the {\iid} hypothesis, arbitrary weighting schemes applied to the data should yield statistically indistinguishable results. 
To operationalize this intuition for complex, potentially non-Euclidean data, we introduce a kernel function to obtain numerical evaluation that can extract high-dimensional features. 
As a consequence, we construct many incomplete U-statistics across strategically designed weighting regimes, which are collected to form a new type of U-process. By quantifying the discrepancy between these incomplete U-statistics, our test statistic essentially measures the cost of being non-\iid. 
This approach transforms the abstract {\iid} testing problem into comparison of weighted means, creating a sensitive detection for hidden structural breaks or dependence patterns. 

A key innovation lies in our treatment of H\'{a}jek projections of the newly introduced U-processes, which leads to an easy-to-implement bootstrap procedure.  
By approximately decomposing the U-process into independent summands, we derive its asymptotic normality while controlling higher-order degenerate components. 
This motivates a jackknife multiplier bootstrap procedure that also serves as a useful addition to the literature \citep{gombay2002rates,chen2018gaussian,chen2020jackknife,han2022multiplier}. 
Accordingly, we reject the {\iid} hypothesis $H_0$ at significance level $\alpha \in (0,1)$ if our test statistic is larger than a data-driven critical value $c_n(\alpha)$ that is easy to compute with external random variables. 

The main contributions of this work are twofold, summarized as follows. 
First, regarding the fundamental problem of checking the {\iid} assumption, we introduce a testing method against general alternatives for random objects. 
To the best of our knowledge, this is the first attempt with such generality and can serve as an important data examination before applying various statistical procedures. 
An easy-to-implement jackknife multiplier bootstrap is tailored to the test statistic, whose desirable performance will be demonstrated in the rest of this paper. 
Second, we have investigated theoretical and numerical properties of the proposed approach. 
The limit theorem for the proposed U-processes, a basic ingredient of our proposed test, is established and gives rise to a new class of Gaussian processes that helps detection of a broad variety of violations of the {\iid} property. 
Under some mild moment conditions, the rates of convergence of the test statistic and its bootstrap are derived in terms of non-asymptotic bounds on Gaussian coupling, which further provides theoretical guarantees on the validity and consistency. 
We examine the versatility of the proposed {\iid} test through simulated and real data examples, comparing with existing methods. 

The rest of the article is organized as follows.
In Section~\ref{sec:method}, we present the construction of our test statistic together with a bootstrap procedure for approximating the null distribution. 
In Section~\ref{sec:theory}, we establish the theoretical results for the proposed {\iid} test, including a local power analysis under a data generation mechanism that incorporates clustered dependencies and sequential distributional changes. 
Simulation studies and real data applications are carried out in Section~\ref{sec:numeric}, validating our method across various data types. 
We collect auxiliary theoretical results and the proofs of theorems, corollaries and technical lemmas in the Supplementary Material.

\section{Proposed Methodology}
\label{sec:method}
In this section, we formalize the core principle underlying our test that violations of {\iid} hypothesis systematically perturb the equilibrium among alternative weighted characterizations of the data. The testing procedure will be derived from a measure of the discrepancies between weighted means of data, together with a bootstrap method for approximating the null distribution of the test statistic. 

\subsection{Test statistic motivated from an off-diagonal U-process}
To begin with, we construct a stochastic process sensitive to both local dependencies and global distributional shifts by evaluating off-diagonal interactions across subject indices. 
Given a symmetric kernel $h : \mathcal{X}\times\mathcal{X} \to \R$ that quantifies pairwise interactions, we define the U-process 
\begin{equation}\label{eq:u-proc}
U_n(t) = (n^2-n)^{-1} \sum_{0<|i-j|\le nt} h(X_i,X_j) ,\quad t\in[0,1] .
\end{equation}
Here the summation excludes diagonal terms and progressively incorporates pairs within expanding index-distance windows, creating a quasi-linear filtration that preserves data adequacy. 
Some choices for $h$ include the characteristic kernel when $\mathcal{X}$ is a reproducing kernel Hilbert space and transformations of the distance function when $\mathcal{X}$ is a metric space. This is inspired by the adaptivity and generality of kernel methods \citep{muandet2017kernel} and metric statistics \citep{dubey2024metric,wang2024nonparametric}. 
The off-diagonal sequential structure of $U_n$ provides dual diagnostic capabilities for the {\iid} assumption. 
Indeed, given the same marginal distribution of $(X_i,X_j)$, the magnitude of $U_n(t)$ is affected by the dependencies between $X_i$'s, since the entire sample $X_{1:n}$ is used. For example, the existence of clustering within $X_{1:n}$ could lead to variance inflation in $U_n(t)$. 
Besides, $U_n(t)$ gives rise to different averages of $h(X_i,X_j)$ as $t$ varies, and the pattern reveals possible distributional nonstationarity of the sequence $X_{1:n}$ in the independent case. 

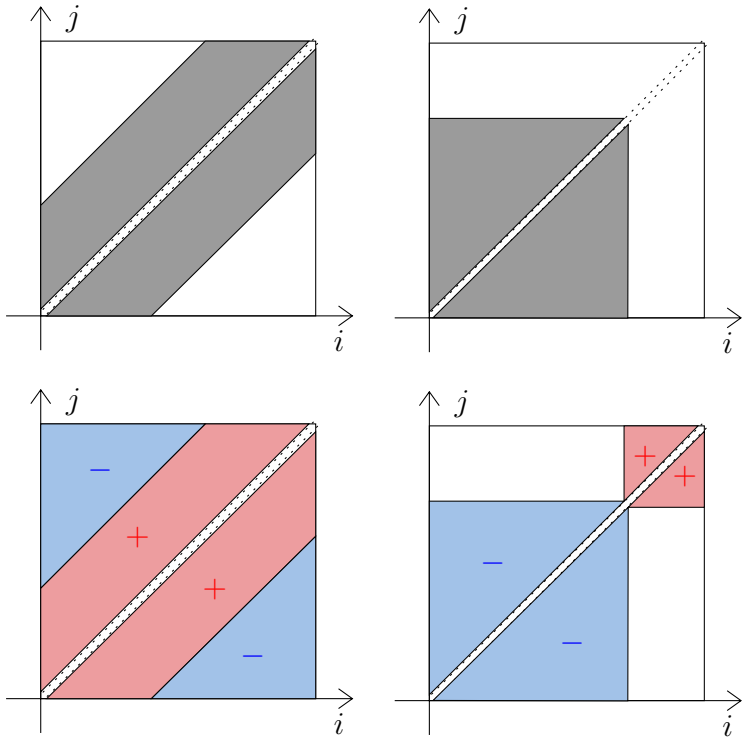
\begin{figure}[!ht]
    \centering
\begin{tikzpicture}[x=0.75pt,y=0.75pt,yscale=-1,xscale=1]

\draw  (50,213.88) -- (224.67,213.88)(67.47,58) -- (67.47,231.2) (217.67,208.88) -- (224.67,213.88) -- (217.67,218.88) (62.47,65) -- (67.47,58) -- (72.47,65)  ;
\draw  [dash pattern={on 0.84pt off 2.51pt}]  (207.21,76.26) -- (68.53,214.94)(205.09,74.14) -- (66.41,212.82) ;
\draw   (67.47,75.2) -- (206.15,75.2) -- (206.15,213.88) -- (67.47,213.88) -- cycle ;
\draw  [fill={rgb, 255:red, 155; green, 155; blue, 155 }  ,fill opacity=1 ] (150.43,75.2) -- (202.74,75.2) -- (67.47,210.4) -- (67.47,158.09) -- cycle ;
\draw  [fill={rgb, 255:red, 155; green, 155; blue, 155 }  ,fill opacity=1 ] (123.29,213.88) -- (70.48,213.88) -- (206.15,79.14) -- (206.15,131.96) -- cycle ;

\draw  (246,214.88) -- (420.67,214.88)(263.47,59) -- (263.47,232.2) (413.67,209.88) -- (420.67,214.88) -- (413.67,219.88) (258.47,66) -- (263.47,59) -- (268.47,66)  ;
\draw  [dash pattern={on 0.84pt off 2.51pt}]  (403.21,77.26) -- (264.53,215.94)(401.09,75.14) -- (262.41,213.82) ;
\draw   (263.47,76.2) -- (402.15,76.2) -- (402.15,214.88) -- (263.47,214.88) -- cycle ;
\draw  [fill={rgb, 255:red, 155; green, 155; blue, 155 }  ,fill opacity=1 ] (263.47,211.88) -- (361.67,114.2) -- (263.47,114.2) -- cycle ;
\draw  [fill={rgb, 255:red, 155; green, 155; blue, 155 }  ,fill opacity=1 ] (363.77,117.3) -- (265.36,214.78) -- (363.56,214.98) -- cycle ;

\draw (214,220) node [anchor=north west][inner sep=0.75pt]   [align=left] {$i$};
\draw (79,56) node [anchor=north west][inner sep=0.75pt]   [align=left] {$j$};
\draw (410,221) node [anchor=north west][inner sep=0.75pt]   [align=left] {$i$};
\draw (275,57) node [anchor=north west][inner sep=0.75pt]   [align=left] {$j$};
\end{tikzpicture}\\[2ex]
\begin{tikzpicture}[x=0.75pt,y=0.75pt,yscale=-1,xscale=1]

\draw  (50,213.88) -- (224.67,213.88)(67.47,58) -- (67.47,231.2) (217.67,208.88) -- (224.67,213.88) -- (217.67,218.88) (62.47,65) -- (67.47,58) -- (72.47,65)  ;
\draw  [dash pattern={on 0.84pt off 2.51pt}]  (207.21,76.26) -- (68.53,214.94)(205.09,74.14) -- (66.41,212.82) ;
\draw   (67.47,75.2) -- (206.15,75.2) -- (206.15,213.88) -- (67.47,213.88) -- cycle ;
\draw  [fill={rgb, 255:red, 237; green, 157; blue, 159 }  ,fill opacity=1 ] (150.43,75.2) -- (202.74,75.2) -- (67.47,210.4) -- (67.47,158.09) -- cycle ;
\draw  [fill={rgb, 255:red, 237; green, 157; blue, 159 }  ,fill opacity=1 ] (123.29,213.88) -- (70.48,213.88) -- (206.15,79.14) -- (206.15,131.96) -- cycle ;
\draw  [fill={rgb, 255:red, 162; green, 194; blue, 232 }  ,fill opacity=1 ] (67.47,75.2) -- (67.47,158.09) -- (150.43,75.2) -- cycle ;
\draw  [fill={rgb, 255:red, 162; green, 194; blue, 232 }  ,fill opacity=1 ] (123.29,213.88) -- (206.15,213.88) -- (206.15,131.96) -- cycle ;

\draw  (246,214.88) -- (420.67,214.88)(263.47,59) -- (263.47,232.2) (413.67,209.88) -- (420.67,214.88) -- (413.67,219.88) (258.47,66) -- (263.47,59) -- (268.47,66)  ;
\draw  [dash pattern={on 0.84pt off 2.51pt}]  (403.21,77.26) -- (264.53,215.94)(401.09,75.14) -- (262.41,213.82) ;
\draw   (263.47,76.2) -- (402.15,76.2) -- (402.15,214.88) -- (263.47,214.88) -- cycle ;
\draw  [fill={rgb, 255:red, 162; green, 194; blue, 232 }  ,fill opacity=1 ] (263.47,211.88) -- (361.67,114.2) -- (263.47,114.2) -- cycle ;
\draw  [fill={rgb, 255:red, 162; green, 194; blue, 232 }  ,fill opacity=1 ] (363.77,117.3) -- (265.36,214.78) -- (363.56,214.98) -- cycle ;
\draw  [fill={rgb, 255:red, 237; green, 157; blue, 159 }  ,fill opacity=1 ] (361.67,76.2) -- (361.67,114.2) -- (399,76.2) -- cycle ;
\draw  [fill={rgb, 255:red, 237; green, 157; blue, 159 }  ,fill opacity=1 ] (363.77,117.3) -- (402.15,117.3) -- (402.15,79) -- cycle ;

\draw (214,220) node [anchor=north west][inner sep=0.75pt]   [align=left] {$i$};
\draw (79,56) node [anchor=north west][inner sep=0.75pt]   [align=left] {$j$};
\draw (410,221) node [anchor=north west][inner sep=0.75pt]   [align=left] {$i$};
\draw (275,57) node [anchor=north west][inner sep=0.75pt]   [align=left] {$j$};

\draw (109,126) node [anchor=north west][inner sep=0.75pt]   [align=left] {\color{red}$+$};
\draw (148,152) node [anchor=north west][inner sep=0.75pt]   [align=left] {\color{red}$+$};
\draw (365,85) node [anchor=north west][inner sep=0.75pt]   [align=left] {\color{red}$+$};
\draw (385,95) node [anchor=north west][inner sep=0.75pt]   [align=left] {\color{red}$+$};
\draw (90,92) node [anchor=north west][inner sep=0.75pt]   [align=left] {\color{blue}$-$};
\draw (167,186) node [anchor=north west][inner sep=0.75pt]   [align=left] {\color{blue}$-$};
\draw (288,139) node [anchor=north west][inner sep=0.75pt]   [align=left] {\color{blue}$-$};
\draw (328,179) node [anchor=north west][inner sep=0.75pt]   [align=left] {\color{blue}$-$};
\end{tikzpicture}
\caption{Indexes of data pairs in off-diagonal sequential U-processes (Left) and sequential U-processes (Right). The two panels on the top correspond to the original versions, while the two panels on the bottom indicate those used for constructing test statistics.}
    \label{fig:idx}
\end{figure}

\begin{remark}
The proposed off-diagonal sequential U-process \eqref{eq:u-proc} exhibits a methodological departure from previously studied U-process frameworks in both construction and purpose. 
A function-indexed U-process is a natural analogue of empirical processes and has found numerous applications in point estimation \citep{arcones1993limit,arcones1994estimators}. 
A sequential U-process is obtained by progressively constructing U-statistics based on a portion of the sample, which facilitates subgraph counting \citep{dobler2022functional} and change-point detection \citep{gombay2002rates,kirch2022sequential}. 
The distinctive difference between off-diagonal sequential U-processes and sequential U-processes lies in how data pairs are selected, as illustrated by the top two panels in Figure~\ref{fig:idx}. The unconventional sampling mechanism within \eqref{eq:u-proc} positions it as a specialized tool for {\iid} testing.
\end{remark}

To test the hypothesis $H_0$ of {\iid} assumption, we examine the stochastic fluctuations of the off-diagonal sequential U-process defined in \eqref{eq:u-proc}. 
Specifically, our proposed test statistic is given by 
\begin{equation}\label{eq:test}
T_n = n^{1/2}\norm{U_n^\diamond}_\infty ,
\end{equation}
where $\norm{\cdot}_\infty$ denotes the supremum norm over the space $L^\infty[0,1]$ of essentially bounded functions, 
\begin{equation}\label{eq:u0-proc}
U_n^\diamond(t) = U_n(t) - u_n(t)U_n(1) ,     
\end{equation}
\begin{equation}\label{eq:u_n}
u_n(t) = (n^2-n)^{-1} \lfloor nt \rfloor(2n-\lfloor nt \rfloor-1) , 
\end{equation}
with $\lfloor d \rfloor$ representing the integer part (floor function) of $d \in \R$. 
The function $u_n$ is obtained precisely as $U_n$ with the constant kernel $h \equiv 1$, so $U_n^\diamond$ reflects the empirical centralization of $U_n$. 
Since the law of large numbers could apply to $U_n(t)$ for a fixed $t$ when $H_0$ holds, any shift in the distributions of the $X_i$'s will be embodied in $U_n^\diamond(t)$. 
Moreover, the construction \eqref{eq:u0-proc} involves evaluating all off-diagonal pairs $(X_i,X_j)$, which enhances its sensitivity to correlation, compared to change-point detection methods based on the difference between two sequential U-processes; see the bottom two panels in Fig.~\ref{fig:idx} for better intuition. 

\subsection{Multiplier bootstrap via H\'{a}jek projection}
We shall utilize the H\'{a}jek projection to understand the asymptotic behavior of the proposed U-processes, which leads to the Gaussian approximation and a pertinent bootstrap procedure that facilitates the implementation of our test. 
The projection method \citep{hajek1968asymptotic} is a powerful device to uncover the underlying probabilistic structure of a statistic. 
Generally, given independent random elements $Y_1,\dots,Y_L$, a random variable $R$ tends to be approximated by $\E(R) + \sum_{k=1}^{L} \varPi_{k}(R)$, where $\varPi_{k}(\cdot) = \E(\cdot\mid Y_k) - \E(\cdot)$. 
We apply it to the off-diagonal sequential U-process \eqref{eq:u-proc} and, under $H_0$ together with the assumption that the second moment of $h(X_1,X_2)$ is finite, simple counting yields 
\[\begin{aligned}
\E\{U_n(t)\mid X_i\} - \E\{U_n(t)\} 
&= (n^2-n)^{-1} \cdot 2 \sum_{j:0<|i-j|\le nt} [\E\{h(X_i,X_j)\mid X_i\} - \E\{h(X_i,X_j)\}] \\
&=: \nu_{ni}(t) h_1^\diamond(X_i) .
\end{aligned}\]
Here $\nu_{ni}(t)$ measures the influence of $X_i$ on $U_n(t)$, defined as $(n^2-n)^{-1}$ multiplied by twice the number of $j=1,\dots,n$ such that $0<|i-j|\le nt$, and thus  
\begin{equation}\label{eq:coef_proj-null}
\nu_{ni}(t) = 2 (n^2-n)^{-1} \{\min(\lfloor nt \rfloor, i-1) + \min(\lfloor nt \rfloor, n-i)\} .
\end{equation}
The function $h_1^\diamond(x)$, $x\in\mathcal{X}$, is defined as the centralization of the partial expectation $h_1(x) = \E\{h(X_1,x)\}$ using the total expectation $h_2 = \E\{h(X_1,X_2)\}$, i.e., 
\[ h_1^\diamond(x) = h_1(x) - h_2 = \E\{h(X_1,x)\} - \E\{h(X_1,X_2)\} ,\quad x\in\mathcal{X} .\]
Now the H\'{a}jek projection of $U_n(t)$ is succinctly given by 
\begin{equation}\label{eq:uproj-proc-null}
\check{U}_n(t) = u_n(t)h_2 + \sum_{i=1}^{n} \nu_{ni}(t) h_1^\diamond(X_i) , 
\end{equation}
with $u_n(t)$ defined in \eqref{eq:u_n}. 
Define correspondingly the empirically centered process 
\begin{equation}\label{eq:u0proj-proc-null}
\check{U}_n^\diamond(t) = \check{U}_n(t) - u_n(t)\check{U}_n(1) = \sum_{i=1}^{n} \nu_{ni}^\diamond(t) h_1^\diamond(X_i) ,
\end{equation}
where $\nu_{ni}^\diamond(t) = \nu_{ni}(t) - 2 n^{-1} u_n(t)$. 
These projected processes $\check{U}_n$ and $\check{U}_n^\diamond$ prove asymptotically equivalent to $U_n$ and $U_n^\diamond$ defined in \eqref{eq:u-proc} and \eqref{eq:u0-proc}, respectively, while their reduced complexity substantially streamlines subsequent analysis.
The representation \eqref{eq:u0proj-proc-null}, although obtained under the null hypothesis, reveals partly how our method transforms abstract {\iid} verification into a measurable imbalance in feature-weight covariation, where the kernel features $h_1^\diamond(X_i)$ and the subject-adjusted weights $\nu_{ni}^\diamond(t)$ bring sensitivity to structural changes.

The linear form of the H\'{a}jek projection enables Gaussian approximations critical for inference, and particularly, we propose a jackknife multiplier bootstrap to address computational concerns for our {\iid} test. 
For dealing with high-dimensional settings which are a major focus in statistics in the last two or three decades, bootstrap has received significant attention recently; see \citet{chernozhukov2023high} for a comprehensive review. 
Our proposed test statistic \eqref{eq:test} is asymptotically equivalent to $\check{T}_n = n^{1/2}\norm{\check{U}_n^\diamond}_\infty$, motivating us to bootstrap the process $\check{U}_n^\diamond = \sum_{i=1}^{n} h_1^\diamond(X_i) \nu_{ni}^\diamond$ in $L^\infty[0,1]$. 
For each $i$, since the population-dependent term $h_1^\diamond(X_i)$ is unknown, we employ its jackknife estimate 
\[ \hat{h}_{1i}^\diamond(X_i) = (n-1)^{-1} \sum_{j\ne i} h(X_i,X_j) - U_n(1) .\]
Such substitutes in function-indexed U-processes were investigated by \citet{chen2020jackknife}. 
In order to extract the distribution of $\check{U}_n^\diamond$ under $H_0$, we introduce the following bootstrapped U-process in light of Gaussian approximation: 
\begin{equation}\label{eq:bootstrap}
\check{U}_n^\epsilon = \sum_{i=1}^{n} \epsilon_i \hat{h}_{1i}^\diamond(X_i) \nu_{ni}^\diamond ,
\end{equation}
where $\epsilon_1,\dots,\epsilon_n$ are {\iid} standard normal random variables that are independent of $X_{1:n}$. 
Then the corresponding bootstrap of our test statistic \eqref{eq:test} is 
\begin{equation}\label{eq:test-bootstrap}
\hat{T}_n = n^{1/2}\norm{\check{U}_n^\epsilon}_\infty .
\end{equation}
This induces a data-driven critical value 
\begin{equation}\label{eq:crit_val}
c_n(\alpha) = \inf\{ t\in\R : \Pr(\hat{T}_n \le t \mid X_{1:n}) \ge 1-\alpha \} ,\quad \alpha \in (0,1) ,
\end{equation}
which admits fast computation as follows. 
Denote by $B$ the number of resamples. 
For $b=1,\dots,B$, while keeping $X_{1:n}$ fixed, generate {\iid} standard normal random variables $\epsilon_1^b,\dots,\epsilon_n^b$ and calculate $\hat{T}_n^b$ as $\hat{T}_n$ in \eqref{eq:test-bootstrap}. 
The critical value $c_n(\alpha)$ is approximately set to be the $(1-\alpha)$th sample quantile of $\{\hat{T}_n^1,\dots,\hat{T}_n^B\}$. 
Equivalently speaking, the $p$-value of the proposed test is approximated by \[ 1 - B^{-1} \sum_{b=1}^{B}\1\{T_n\ge\hat{T}_n^b\} ,\]
where $\1\{\cdot\}$ denotes the indicator function.

\section{Theoretical Guarantees}
\label{sec:theory}
In this section, we explore the theoretical properties of the proposed method, aimed at the validity and consistency of our {\iid} test. 
To simplify the presentation, we introduce the notation of inequalities up to constant factors. 
In what follows, denote $\theta \lesssim \phi$ or $\theta = \order{\phi}$ for real quantities $\theta,\phi$ when there exists a numerical constant $C>0$ such that $\abs{\theta} \le C\phi$. 
For example, the weighting functions \eqref{eq:u_n} and \eqref{eq:coef_proj-null} satisfy that $\abs{u_n(t)-u_n(s)} \lesssim \abs{t-s}+n^{-1}$ and $\abs{\nu_{ni}(t)-\nu_{ni}(s)} \lesssim n^{-1}(\abs{t-s}+n^{-1})$, where the former can be derived from the latter since $u_n = 2^{-1} \sum_{i=1}^{n} \nu_{ni}$. 

\subsection{Validity of the bootstrap procedure}
In order to justify the proposed test, the critical value defined in \eqref{eq:crit_val} will prove valid for controlling the type~I error. 

We first show that the H\'{a}jek projection used in Section~\ref{sec:method} brings small perturbations to the proposed U-processes, underpinning the bootstrap procedure based on \eqref{eq:u0proj-proc-null}. 
The following Theorem~\ref{thm:proj-null} characterizes the uniform approximation error for the projection. 
\begin{theorem}\label{thm:proj-null}
Let $H_0$ hold. 
Recall $U_n(t),U_n^\diamond(t),\check{U}_n(t),\check{U}_n^\diamond(t)$ given in \eqref{eq:u-proc},\eqref{eq:u0-proc},\eqref{eq:uproj-proc-null},\eqref{eq:u0proj-proc-null}. 
If 
\[ \E\{h(X_1,X_2)^2\} \lesssim \sigma^2 \] 
for some constant $\sigma > 0$ not depending on $n$, 
then we have 
\begin{align*}
\E\big(n\norm{U_n-\check{U}_n}_\infty^2\big) &\le \rho_n , \\
\E\big(n\norm{U_n^\diamond-\check{U}_n^\diamond}_\infty^2\big) &\le \rho_n ,
\end{align*}
where $\rho_n = \order{ \sigma^2 n^{-1/3} }$. 
\end{theorem}
\begin{remark}
Motivated by the classical Glivenko--Cantelli theorem, the proof of Theorem~\ref{thm:proj-null} relies on the sequential property along $[0,1]$. 
The rate $n^{-1/3}$ in Theorem~\ref{thm:proj-null} may be not optimal, but is sufficient for our purpose. That is, we are able to bound the projection error of the proposed test statistic \eqref{eq:test}. 
\end{remark}

Next we establish the Gaussian approximation for the proposed U-processes, justifying our introduction of normally distributed multipliers in \eqref{eq:bootstrap}. 
Recall that a Gaussian process $W$ on an index set $\mathcal{T}$ is a collection of random variables $(W(t))_{t\in\mathcal{T}}$ such that every finite subcollection $(W(t_1),\dots,W(t_m))$ has a multivariate normal distribution. 
Under $H_0$, suppose that 
\begin{equation*}
\Gamma_h(s,t) = \lim_{n\to\infty} \Cov\{n^{1/2}\check{U}_n(s),n^{1/2}\check{U}_n(t)\} = \lim_{n\to\infty} n\sum_{i=1}^{n}\nu_{ni}(s)\nu_{ni}(t) \E\{h_1^\diamond(X_1)^2\} 
\end{equation*}
exists for any $s,t\in[0,1]$, which holds when $\E\{h_1^\diamond(X_1)^2\} = \sigma^2 + \order{n^{-2}}$ by Lemma~\ref{lem:cov} in Appendix~\ref{appn:theory}. 
Let $G$ be a zero-mean Gaussian process on $[0,1]$ with $\Gamma_h$ being its covariance function, and let \[ G^\diamond = G - G(1)u_\infty ,\] where $u_\infty(t) = \lim_{n\to\infty}u_n(t) = t(2-t)$ for $t\in[0,1]$. 
\begin{theorem}\label{thm:clt-null}
Let $H_0$ hold. 
Recall $\check{U}_n(t)$ in \eqref{eq:uproj-proc-null}. 
Assume that 
\[ 
\E\{|h(X_1,X_2)|^3\} \lesssim \sigma^3 \]
for some constant $\sigma > 0$ not depending on $n$. 
Then $n^{1/2}(\check{U}_n - h_2 u_n)$ converges in distribution to $G$ in $L^\infty[0,1]$. 
\end{theorem}
Theorem~\ref{thm:clt-null} is affirmed by verifying a uniform central limit theorem. 
The assumption of a finite third moment enables weak convergence to the prescribed Gaussian process, which can be regarded as a natural extension of Lyapunov's condition for the central limit theorem. 
Since the projections are good approximations by Theorem~\ref{thm:proj-null}, the convergence can be extended to the original U-processes. 
\begin{corollary}\label{cor:clt-null}
Let $H_0$ hold. 
Recall $U_n(t),U_n^\diamond(t),\check{U}_n^\diamond(t)$ in \eqref{eq:u-proc},\eqref{eq:u0-proc},\eqref{eq:u0proj-proc-null}. 
Under the assumptions of Theorem~\ref{thm:clt-null}, 
\begin{itemize}
    \item $n^{1/2}(U_n - h_2 u_n)$ converges in distribution to $G$ in $L^\infty[0,1]$, 
    \item both $n^{1/2} U_n^\diamond$ and $n^{1/2} \check{U}_n^\diamond$ converge in distribution to $G^\diamond$ in $L^\infty[0,1]$. 
\end{itemize}
\end{corollary}

Corollary~\ref{cor:clt-null} implies that under the {\iid} assumption, the sampling distributions of the proposed test statistic and its bootstrapped version (conditional on the sample) are both close to the distribution of $\norm{G^\diamond}_\infty$, supporting the choice of the critical value $c_n(\alpha)$ defined in \eqref{eq:crit_val}. 
More precisely, we derive a high-probability bound on their Kolmogorov distance, shown in the following Theorem~\ref{thm:dist-null}. 
\begin{theorem}\label{thm:dist-null}
Recall $T_n$ and $\hat{T}_n$ defined in \eqref{eq:test} and \eqref{eq:test-bootstrap}. 
Under $H_0$, if \[ \E\{h_1^\diamond(X_1)^2\} = \sigma^2 + \order{n^{-2}} \qand \E\{h_1^\diamond(X_1)^4\} \lesssim \sigma^4 \] for some constant $\sigma > 0$, then there exists some 
\[ \omega_{nc} = \order{c^{-3/2} n^{-1/10} \log^{3/4}n} ,\quad c\in(0,1) ,\]
such that with probability at least $1-c$, 
\[ \sup_{t\in\R}\lvert \Pr(T_n \le t) - \Pr(\hat{T}_n \le t \mid X_{1:n}) \rvert \le \omega_{nc} .\]
\end{theorem}
Theorem~\ref{thm:dist-null} ensures that our proposed test is valid in the sense that the size, or type~I error rate, is controlled up to a sufficiently small term: 
\[ \abs{\Pr\{T_n > c_n(\alpha)\} - \alpha} \le \omega_{nc} .\]
The bound of $\omega_{nc}$ in terms of $n$ reflects the trade-off between Gaussian coupling for $T_n$ and $\hat{T}_n$, while taking into account the projection error characterized by Theorem~\ref{thm:proj-null}.

\subsection{Local power analysis}
Now we investigate the local power of the proposed {\iid} test. 
To facilitate theoretical analysis and for specificity, consider the sample $X_{1:n}$ generated as 
\[ H_1 : X_{i} = Y_{k} \qfor i \in \mathcal{I}_{k} ,\]
where $Y_{1},\dots,Y_{L}$ are independent random elements, and $\mathcal{I}_{1},\dots,\mathcal{I}_{L}$ are disjoint sets that constitute a partition of $\{1,\dots,n\}$. 
Such a data generation mechanism incorporates clustered dependencies and sequential distributional changes, while the {\iid} assumption $H_0$ becomes a degenerate case. 
In particular, there are two special cases of clustering and change-point, respectively, which we specify as follows: 
\begin{itemize}
    \item (Clustering). $H_1^{\mathrm{cl}} : X_i = Y_{\lfloor (i-1)/m + 1 \rfloor}$, where $Y_1,Y_2,\dots$ are {\iid} and $m$ is a fixed positive integer standing for the cluster size. Here $\mathcal{I}_{k} = \{i : (k-1)m < i \le km\}$.
    \item (Change-point). $H_1^{\mathrm{cp}} : X_1,\dots,X_n$ are independent random elements such that $X_i$ is distributed as $Y^< \1\{i\le n\tau\} + Y^> \1\{i>n\tau\}$, where $Y^< , Y^>$ are independent and $\tau\in(0,1)$ is a fixed number locating the change-point. In this case, $\mathcal{I}_{k} = \{k\}$.
\end{itemize}

Then the off-diagonal sequential U-process \eqref{eq:u-proc} can be rewritten as 
\begin{equation}\label{eq:u-proc_re}
\begin{aligned}
U_n(t) 
&= (n^2-n)^{-1} \sum_{k,\ell=1}^{L} \sum_{i\in\mathcal{I}_k} \sum_{j\in\mathcal{I}_\ell} \1\{0<|i-j|\le nt\} h(Y_k,Y_\ell) \\
&= \sum_{k=1}^{L} u_{nkk}(t) h(Y_k,Y_k) + 2 \sum_{1\le k < \ell \le L} u_{nk\ell}(t) h(Y_k,Y_\ell) 
=: U_{n1}(t) + U_{n2}(t) ,
\end{aligned}
\end{equation}
where $u_{nk\ell}(t)$ accounts for the weight of $(Y_k,Y_\ell)$ in $U_n(t)$, given by 
\[ u_{nk\ell}(t) = (n^2-n)^{-1} \sum_{i\in\mathcal{I}_k}\sum_{j\in\mathcal{I}_\ell} \1\{0<|i-j|\le nt\} ,\quad k,\ell=1,\dots,L .\]
We assume that the cardinality $|\mathcal{I}_{k}|$ is bounded by a constant $m$ not depending on $n$, so $u_{nk\ell}(t) \le (n^2-n)^{-1} |\mathcal{I}_{k}| \cdot |\mathcal{I}_{\ell}| \le 2 m^{2} n^{-2}$. This implies that each pair $(Y_k,Y_\ell)$ has a sufficiently small influence on $U_n(t)$, and also that the number of independent random elements is large enough as $L \ge n/m$. 
As a consequence, the variation of $U_n(t)$ reflects the joint distribution of $X_{1:n}$, where the magnitude of $u_{nk\ell}(t)$ exhibits the size of the clusters, and the changes of $u_{nk\ell}(t)$ with respect to $t$ highlight the individual role of $(Y_k,Y_\ell)$. 

We go through with the H\'{a}jek projection to unveil the probabilistic structure of \eqref{eq:u-proc_re}, and the subtleties of construction become clearer. 
For generality, we extend the projection representations \eqref{eq:uproj-proc-null} and \eqref{eq:u0proj-proc-null} in the setting of $H_1$ with all $h(Y_k,Y_\ell)$ having finite second moments. 
Let  
\begin{equation}\label{eq:uproj-proc}
\check{U}_n(t) = \mu_n(t) + \sum_{k=1}^{L}\varPi_{k}\{U_n(t)\} ,
\end{equation}
\begin{equation}\label{eq:u0proj-proc}
\check{U}_n^\diamond(t) = \check{U}_n(t) - u_n(t)\check{U}_n(1) = \mu_n^\diamond(t) + \sum_{k=1}^{L}\varPi_{k}\{U_n^\diamond(t)\} ,
\end{equation}
where $\varPi_{k}(\cdot) = \E(\cdot\mid Y_k) - \E(\cdot)$ is the H\'{a}jek projection operator, $\mu_n(t) = \E\{U_n(t)\}$ and $\mu_n^\diamond(t) = \E\{U_n^\diamond(t)\} = \mu_n(t) - u_n(t)\mu_n(1)$. 
Note that simple calculation leads to 
\[\begin{aligned}
\varPi_{k}\{U_n(t)\} &= \varPi_{k}\{U_{n1}(t)\} + \varPi_{k}\{U_{n2}(t)\} \\
&= u_{nkk}(t) \varPi_{k}\{h(Y_k,Y_k)\} + 2 \sum_{\ell:\ell\ne k} u_{nk\ell}(t) \varPi_{k}\{h(Y_k,Y_\ell)\} ,
\end{aligned}\]
\[\begin{aligned}
\varPi_{k}\{U_n^\diamond(t)\} &= \varPi_{k}\{U_n(t)\} - u_n(t) \varPi_{k}\{U_n(1)\} \\
&= u_{nkk}^\diamond(t) \varPi_{k}\{h(Y_k,Y_k)\} + 2 \sum_{\ell:\ell\ne k} u_{nk\ell}^\diamond(t) \varPi_{k}\{h(Y_k,Y_\ell)\} ,
\end{aligned}\]
where $u_{nk\ell}^\diamond(t) = u_{nk\ell}(t) - u_n(t) u_{nk\ell}(1)$. 
See Appendix~\ref{appn:theory} for better understanding of \eqref{eq:uproj-proc} and \eqref{eq:u0proj-proc}. 
To characterize the variability, denote \[ D_k = \max_{1 \le \ell \le L} \max[\abs{\varPi_{k}\{h_+(Y_k,Y_\ell)\}}, \abs{\varPi_{k}\{h_-(Y_k,Y_\ell)\}}] ,\]
where $h_+$ and $h_-$ are the positive and negative parts of $h$, respectively. 

Under $H_1$, Gaussian approximation still plays an important role. 
In what follows, we generalize the previously defined Gaussian processes with a slight abuse of notation. 
Suppose that 
\begin{equation}\label{eq:cov}
\Gamma_h(s,t) = \lim_{n\to\infty} \Cov\{n^{1/2}\check{U}_n(s),n^{1/2}\check{U}_n(t)\} = \lim_{n\to\infty} n \sum_{k=1}^{L} \Cov[\varPi_{k}\{U_n(s)\},\varPi_{k}\{U_n(t)\}]
\end{equation}
exists for any $s,t\in[0,1]$. 
Such convergence is demonstrated in Appendix~\ref{appn:theory}. 
Let $G$ be a zero-mean Gaussian process on $[0,1]$ with $\Gamma_h$ being its covariance function, and let \[ G^\diamond = G - G(1)u_\infty .\]
The proposed test statistic $T_n = n^{1/2}\norm{U_n^\diamond}_\infty$ is approximated by $\norm{n^{1/2}\mu_n^\diamond + G^\diamond}_\infty$, the supremum norm of a Gaussian process with fairly complicated mean and covariance functions, which facilitates detection of departure from the {\iid} assumption. 

Since the critical value \eqref{eq:crit_val} is based on the conditional distribution of the bootstrapped test statistic \eqref{eq:test-bootstrap}, the asymptotic behavior of the bootstrapped U-process \eqref{eq:bootstrap} is intimately relevant. 
We introduce a Gaussian process corresponding to its limiting distribution. 
Let \[ h_1^\diamond(x) = n^{-1}\sum_{i=1}^{n} \E\{h(X_i,x)\} - \E\{U_n(1)\} ,\quad x\in\mathcal{X} ,\]
which extends beyond the {\iid} setting in Section~\ref{sec:method}. 
The integrability of $h_1^\diamond(X_i)$, $i=1,\dots,n$, is easily seen from the fact that 
$\abs{h_1^\diamond(X_i)} \lesssim D_{k(i)} + \max_{1 \le k,\ell \le L} \E\{\abs{h(Y_k,Y_\ell)}\}$, where $k(i)$ is defined by $i \in \mathcal{I}_{k(i)}$. 
In analogy to \eqref{eq:cov}, suppose that 
\[ \tilde{\Gamma}_h^\diamond(s,t) = \lim_{n\to\infty} n \sum_{i=1}^{n} \nu_{ni}^\diamond(s)\nu_{ni}^\diamond(t) \E\{h_1^\diamond(X_i)^2\} \]
exists for any $s,t\in[0,1]$. 
Let $\tilde{G}$ be a zero-mean Gaussian process on $[0,1]$ with covariance function $\tilde{\Gamma}_h^\diamond$, which serves as the limit of \eqref{eq:bootstrap}. 
Then the data-driven critical value \eqref{eq:crit_val} has the underlying limit 
\[ c_*(\alpha) = \inf\{ t\in\R : \Pr(\lVert\tilde{G}\rVert_\infty \le t) \ge 1-\alpha \} .\]

To quantify the rate of convergence of the U-processes \eqref{eq:bootstrap} and \eqref{eq:u0proj-proc}, the approximation errors of covariance functions will be useful. Define 
\[ \tilde{\Delta}_n = \max_{1\le j,k \le n} \abs{n\sum_{i=1}^{n}\nu_{ni}^\diamond(jn^{-1})\nu_{ni}^\diamond(kn^{-1})\E\{h_1^\diamond(X_i)^2\} - \tilde{\Gamma}_h^\diamond(jn^{-1},kn^{-1})} ,\]
\begin{equation*}
\Delta_n = \max_{1\le j,k \le n} \abs{n \sum_{i=1}^{L} \Cov[\varPi_{i}\{U_n^\diamond(jn^{-1})\},\varPi_{i}\{U_n^\diamond(kn^{-1})\}] - \Gamma_h^\diamond(jn^{-1},kn^{-1})} ,
\end{equation*}
where $\Gamma_h^\diamond$ is the covariance function of the Gaussian process $G^\diamond$. 
In view of Lemma~\ref{lem:cov} in Appendix~\ref{appn:theory}, we typically have $\tilde{\Delta}_n = \mathcal{O}\{(\sigma^2+M_1^2) n^{-1}\}$ and $\Delta_n = \order{m^2 \sigma^2 n^{-1}}$, provided that $\max_{1\le k\le L} \E(D_k^2) \lesssim \sigma^2$ and $\max_{1 \le k,\ell \le L} \E\{\abs{h(Y_k,Y_\ell)}\} \lesssim M_1$. 

Finally we obtain the following lower bound on the power of our proposed test. 
\begin{theorem}\label{thm:power}
Let $H_1$ hold. 
Assume that \[ \max_{1 \le k,\ell \le L} \E\{\abs{h(Y_k,Y_\ell)}\} \lesssim M_1 ,\] 
\[ \max_{1\le k < \ell \le L} \Var\{h(Y_k,Y_\ell)\} \lesssim M_2 ,\]
\[ \max_{1\le k\le L} \E(D_k^4) \lesssim \sigma^4 \] 
for some constants $M_1,M_2,\sigma > 0$ not depending on $n$, 
and that $\tilde{\Delta}_n = \mathcal{O}\{(\sigma^2+M_1^2) n^{-1/2}\}$ and $\Delta_n = \order{m^2 \sigma^2 n^{-1/2}}$. 
Then for any $\gamma \in (0,1)$, 
\[ \Pr\{T_n > c_n(\alpha)\} \ge \Pr\big\{\norm{n^{1/2}\mu_n^\diamond + G^\diamond}_\infty > c_*(\alpha-2\gamma) + r_{n\gamma}\big\} - 4\gamma ,\]
where \[\begin{aligned}
r_{n\gamma} = \mathcal{O}\big( &\gamma^{-1/2} n^{-1/6} \{m M_2^{1/2} + M_1 + m \sigma (n^{-1/6} + \gamma^{1/6}\log^{2/3}n)\} \\ &
+ \gamma^{-3} n^{-1/4} \{m\sigma + (\sigma+M_1)\log^{1/4}n + n^{-1/2}m M_2^{1/2}\} \log^{1/2}n \big).
\end{aligned}\]
\end{theorem}
The magnitude of $r_{n\gamma}$ is kind of complicated, combining errors of projection, coupling, and Gaussian comparison, but the crux is that $\lim_{n\to\infty} r_{n\gamma} = 0$. 
Since $\gamma\in(0,1)$ can be chosen arbitrarily small, Theorem~\ref{thm:power} implies that the power is asymptotically given by \[ \Pr\big\{\norm{n^{1/2}\mu_n^\diamond + G^\diamond}_\infty > c_*(\alpha)\big\} ,\] which increases with the deviation of the sample-induced Gaussian process $n^{1/2}\mu_n^\diamond + G^\diamond$ from the bootstrap-induced Gaussian process $\tilde{G}$. 
Specifically, we arrive at the following consistency results for the local alternatives $H_1^{\mathrm{cl}}$ and $H_1^{\mathrm{cp}}$. 
\begin{corollary}\label{cor:power}
Let the assumptions of Theorem~\ref{thm:power} hold.
\begin{itemize}[leftmargin=*]
    \item Under $H_1^{\mathrm{cl}}$, for any $\beta\in(0,1)$, there exists some constant $m_0 > 0$ only depending on $\beta$ such that if $m \ge m_0$, then 
\[ \liminf_{n\to\infty} \Pr\{T_n > c_n(\alpha)\} \ge 1 - \beta .\]
    \item Under $H_1^{\mathrm{cp}}$, for any $\beta\in(0,1)$, there exists some constant $\mu_0 > 0$ only depending on $\beta$ such that if 
    \[ \liminf_{n\to\infty} n^{1/2}\big[|\E\{h(Y^<,Y'^<)\}-\E\{h(Y^<,Y^>)\}|+|\E\{h(Y^>,Y'^>)\}-\E\{h(Y^<,Y^>)\}|\big] \ge \sigma\mu_0 ,\]
    where $(Y'^< , Y'^>)$ is an independent copy of $(Y^< , Y^>)$, then 
\[ \liminf_{n\to\infty} \Pr\{T_n > c_n(\alpha)\} \ge 1 - \beta .\]
\end{itemize}
\end{corollary}
Corollary~\ref{cor:power} shows that the power becomes arbitrarily close to $1$ when the detection threshold is achieved. 
The conditions align with our intuition, i.e., the larger cluster size $m$ under $H_1^{\mathrm{cl}}$ or the larger change magnitude under $H_1^{\mathrm{cp}}$ results in rejection of $H_0$ with higher probability.

\section{Numerical studies}
\label{sec:numeric}
This section illustrates the numerical performance of our proposed {\iid} test. 
We first exhibit its power and versatility under various settings, and then head for applications to air pollutants data, the MNIST data of handwritten digits and two time series of multilayer network data reflecting financial linkage and email activity. 

\subsection{Simulation}
\label{sec:simu}
Frequently used methods for {\iid} testing are often devised against specific alternatives, among which for comparison we consider the change-point test and the white noise test developed by \citet{yu2022robust} and \citet{fokianos2018testing}, respectively. 
Besides, a popular way to detect distribution drift in machine learning is based on PCA reconstruction error \citep{NannyML}. 
To comprehensively understand the size and power of these tests, we introduce six models according to which the sample $X_1,\dots,X_n$ is generated, with $\varepsilon_1,\varepsilon_2,\dots$ being {\iid} standard normal random vectors of dimension $p=5$. 
\begin{itemize}
    \item Mean drift (MD): $X_i = i\mu e + \varepsilon_i$, $i=1,\dots,n$, where $e$ is the vector of all $1$s; 
    \item Variance change-point (VCP): $X_i = (\1\{i\le n/2\} + \sigma\1\{i>n/2\}) \varepsilon_i$, $i=1,\dots,n$; 
    \item Autoregression (AR): $X_i = \varepsilon_i$, $i=1,2$, and $X_i = a (X_{i-1} - X_{i-2}) + \varepsilon_i$, $i=3,\dots,n$; 
    \item Moving average (MA): $X_1 = \varepsilon_1$ and $X_i = \varepsilon_i + b \varepsilon_{i-1}$, $i=2,\dots,n$. 
    \item Mean drift in moving average (MDMA): $X_1 = \mu e + \varepsilon_1$ and \[ X_i = i\mu e + \varepsilon_i + b \varepsilon_{i-1} ,\quad i=2,\dots,n .\]
    \item Variance change-point in moving average (VCPMA): $X_1 = \varepsilon_1$ and \[ X_i = (\1\{i\le n/2\} + \sigma\1\{i>n/2\}) (\varepsilon_i + b \varepsilon_{i-1}) ,\quad i=2,\dots,n . \]
\end{itemize}
Here some parameters, $\mu,\sigma,a,b$, are incorporated to enhance variability. 
The model $\mathcal{M}_0$ that $X_i = \varepsilon_i$ corresponds to the {\iid} case where rejecting $H_0$ implies a type~I error. 

Our {\iid} test, denoted by \textsc{odsup}, is implemented with $h(x,y) = \e^{-\norm{x-y}}$ where $\norm{\cdot}$ is the Euclidean norm. 
In the \textsc{pca} approach by \citet{NannyML}, we take $2$ principal components and divide the sample by the parity of the time order $i$, being even or odd, to generate reference and analysis data. 
Choosing $h^{(1)}(x,y)=(x_j-y_j)_{1\le j\le p}$ and $h^{(2)}(x,y)=(x_j^2-y_j^2)_{1\le j\le p}$, for $x=(x_j)_{1\le j\le p}$ and $y=(y_j)_{1\le j\le p}$, gives rise to two change-point tests following \citet{yu2022robust}, say, \textsc{cp1} and \textsc{cp2}, respectively. 
The white noise test based on auto-distance correlation (resp.\ covariance) is abbreviated as \textsc{adcr} (resp.\ \textsc{adcv}), where we use the bandwidth $\lfloor 3 n^{1/5} \rfloor$ and the Bartlett kernel \citep{fokianos2018testing}. 

We conduct the six tests in the above-defined models with different parameters and sample sizes, given the nominal significance level $\alpha = 5\%$. 
To assess their empirical power, rejection proportions are calculated based on 1000 Monte Carlo replications, as shown in Figure~\ref{fig:power} and Tables~\ref{tab:power}--\ref{tab:pow_cont} in Appendix~\ref{appn:simu}. 
In the model $\mathcal{M}_0$, we see that the empirical sizes of all tests are around the nominal level. 
Regarding power performance under alternatives of the first four models, even though the specifically devised tests may demonstrate advantages within their respective domains, our method \textsc{odsup} consistently maintains its standing as a reasonable choice. 
This is primarily attributable to the comprehensive adaptability inherent in the proposed approach, allowing it to effectively address a diverse range of alternatives. 
On the other hand, \textsc{pca} fails in every scenario, \textsc{cp1} is unable to tackle VCP, AR and MA, \textsc{cp2} shows inadequacies when dealing with MA, and \textsc{adcr} and \textsc{adcv} lose power for VCP. 
Thus, while conceding the superiority of specialized methods in their designated areas, \textsc{odsup} remains reliable and versatile that is useful in various contexts. 
We see that \textsc{odsup} dominates the other tests under some mixed designs. For instance, given a moving average background, mean drift or variance change-point is more challenging to be detected. This results in some cases of MDMA and VCPMA where \textsc{odsup} achieves the highest power. 
This provides empirical evidence that the proposed \textsc{odsup} test is more applicable to complex-structured data, which are common in the real world today. 

\begin{figure}[!ht]
    \centering
\includegraphics[width=0.49\linewidth]{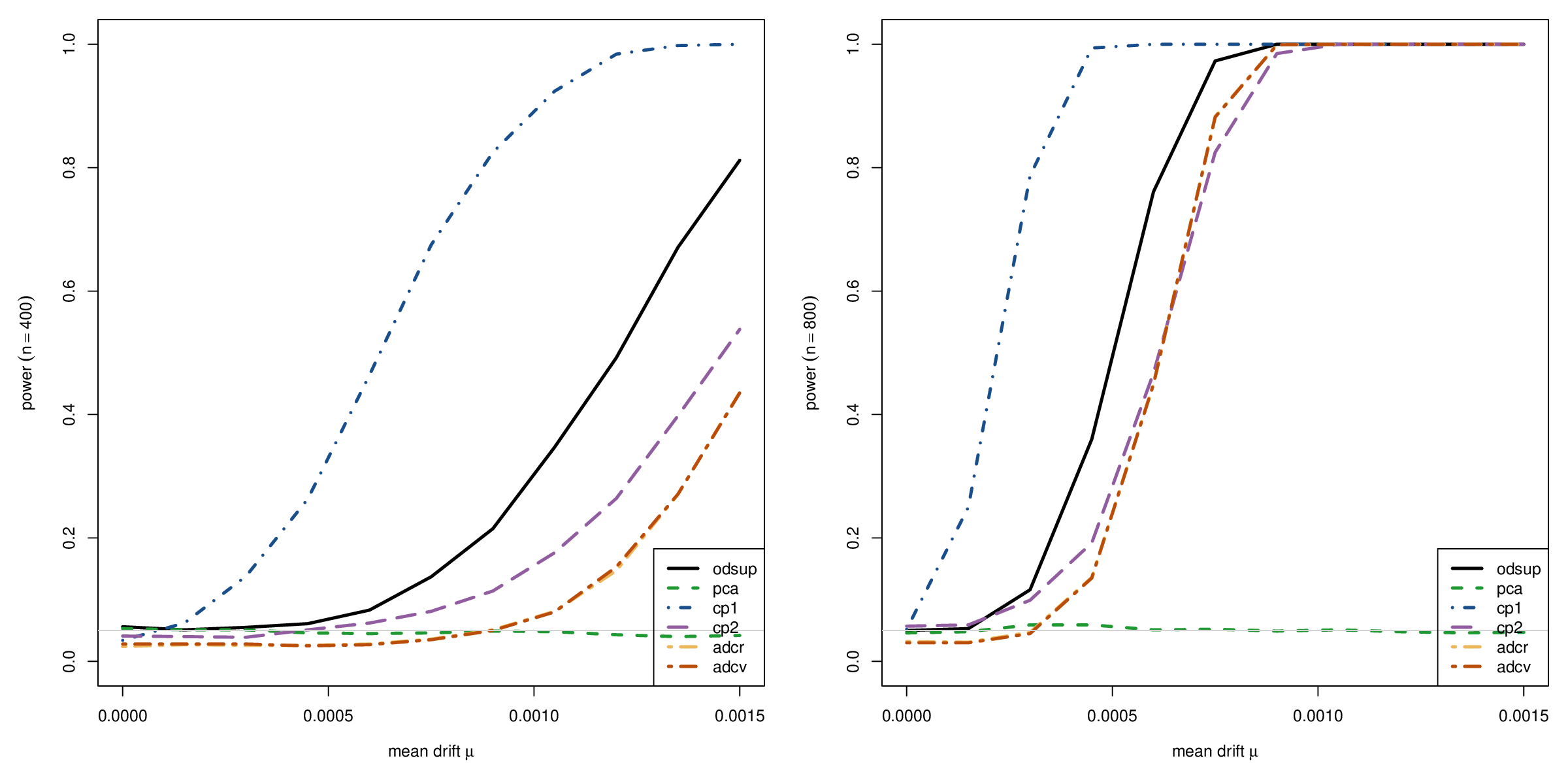}
\includegraphics[width=0.49\linewidth]{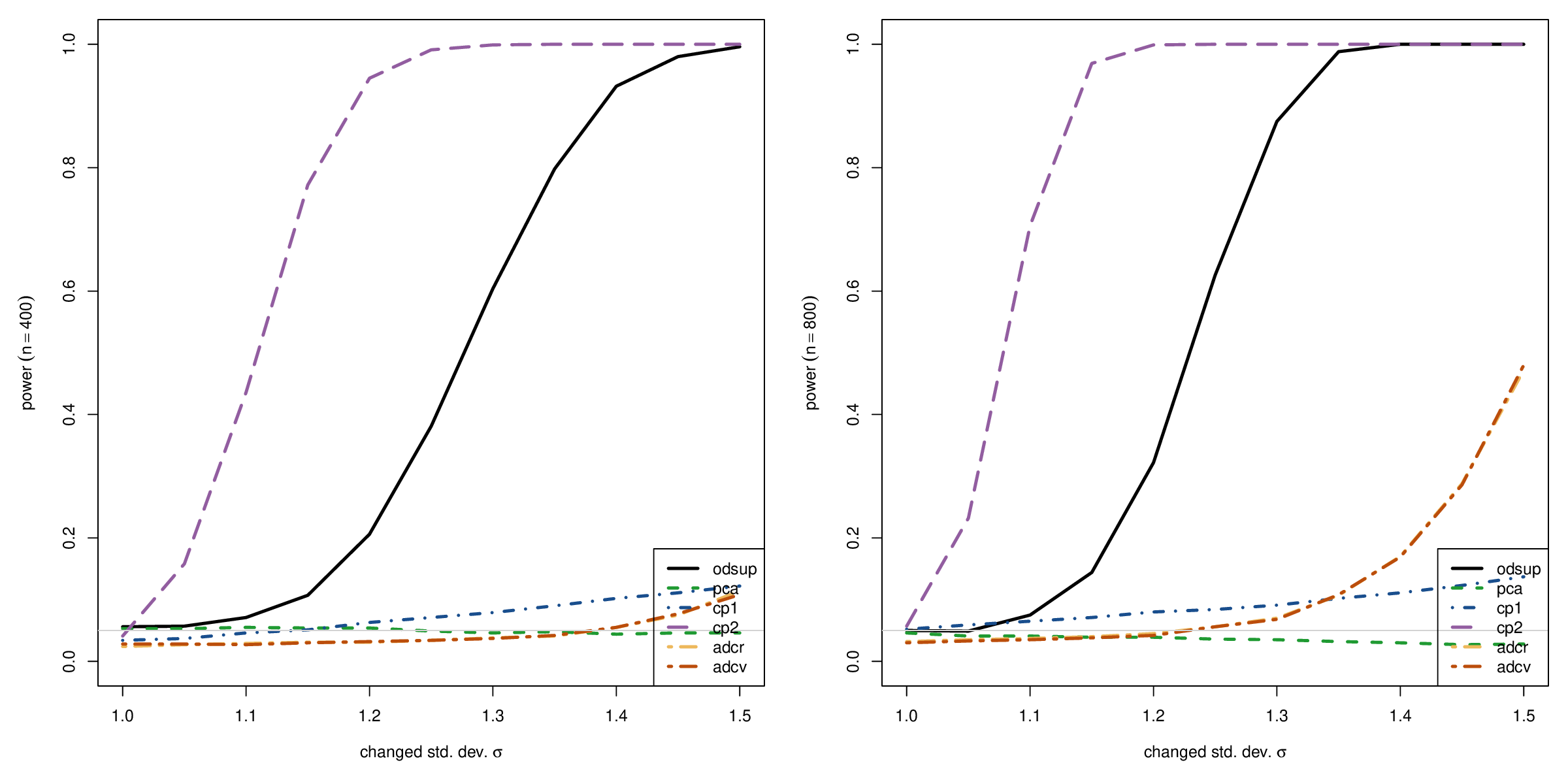}
\includegraphics[width=0.49\linewidth]{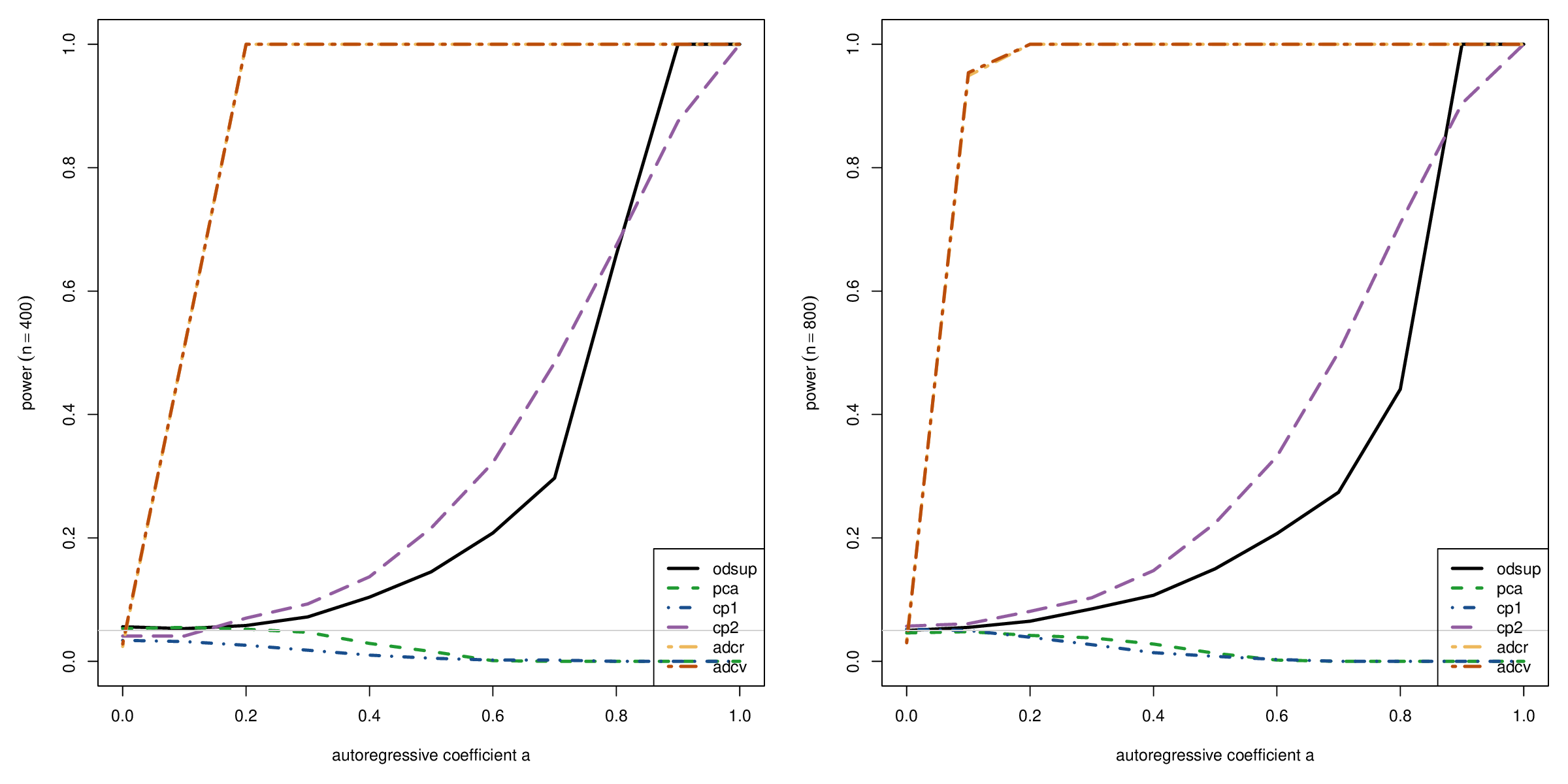}
\includegraphics[width=0.49\linewidth]{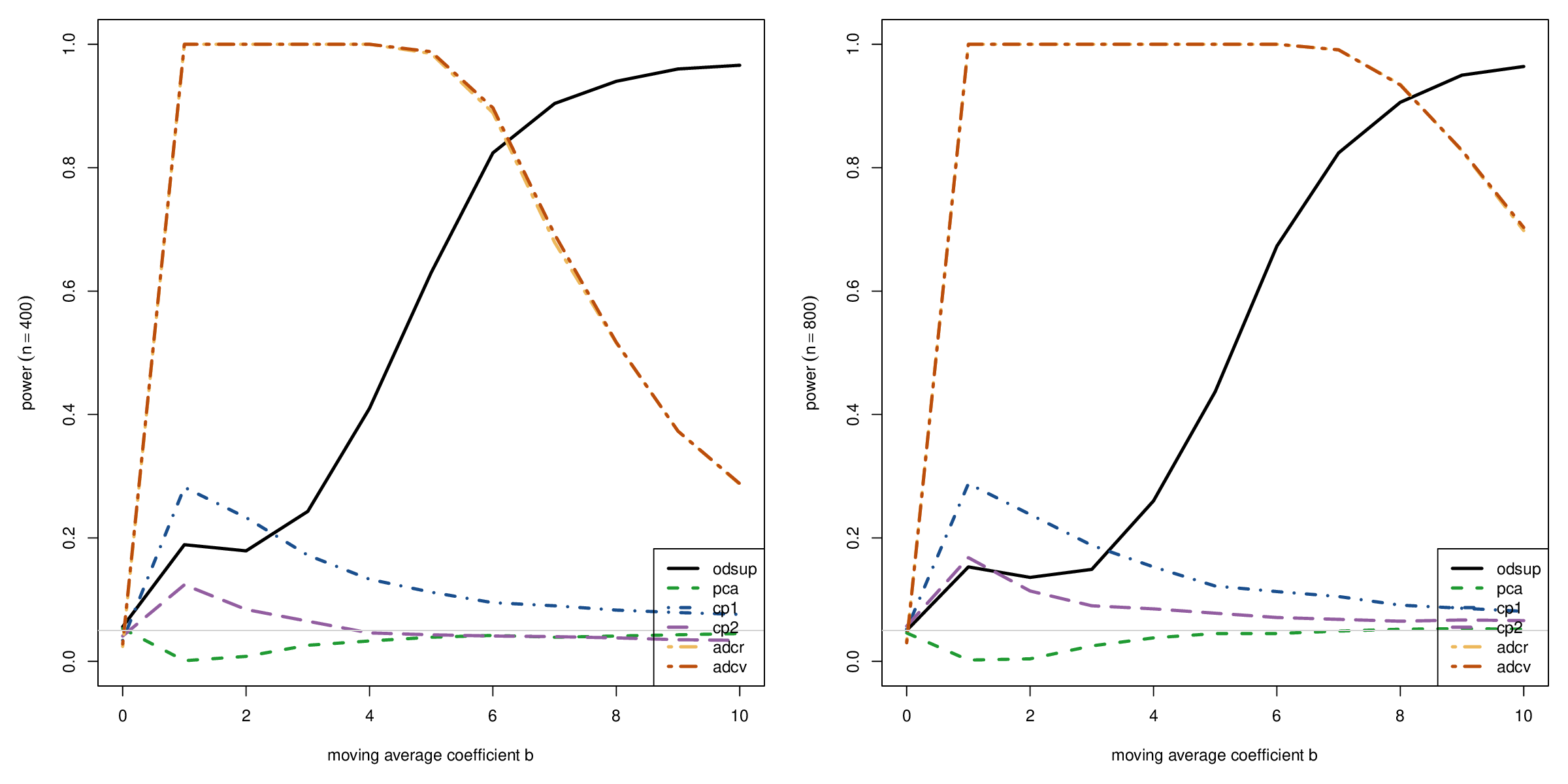}
\includegraphics[width=0.49\linewidth]{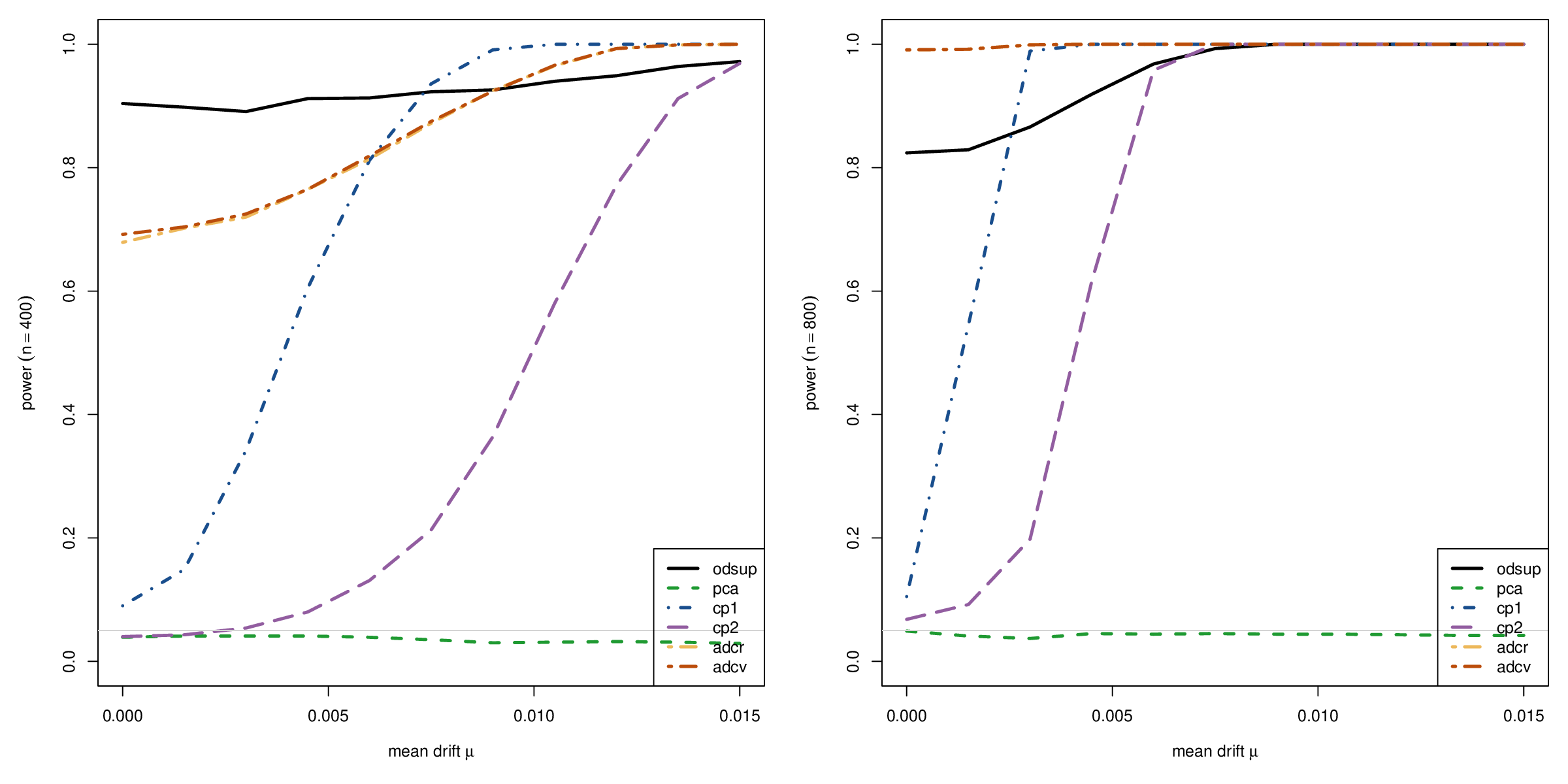}
\includegraphics[width=0.49\linewidth]{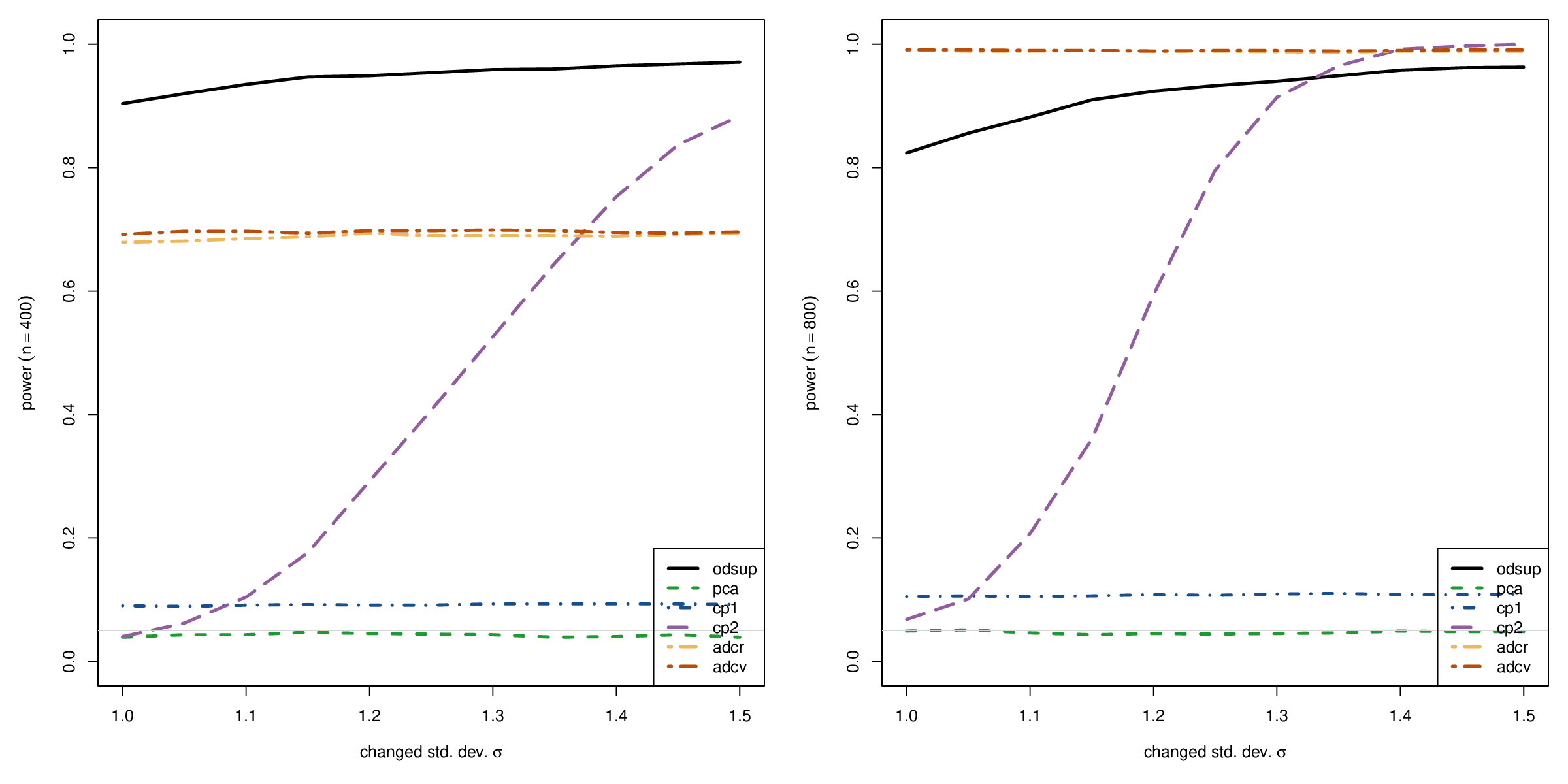}
\includegraphics[width=0.49\linewidth]{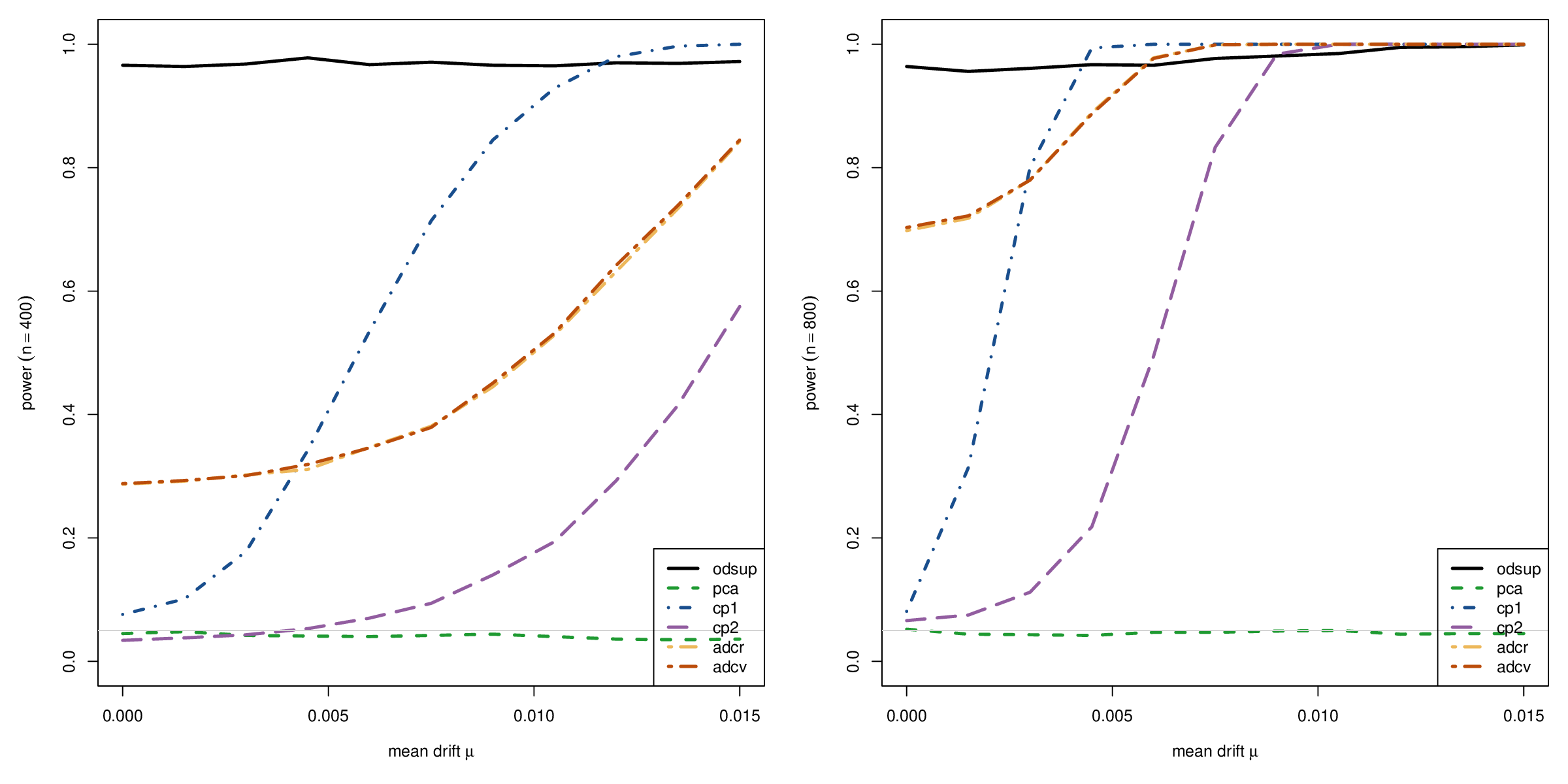}
\includegraphics[width=0.49\linewidth]{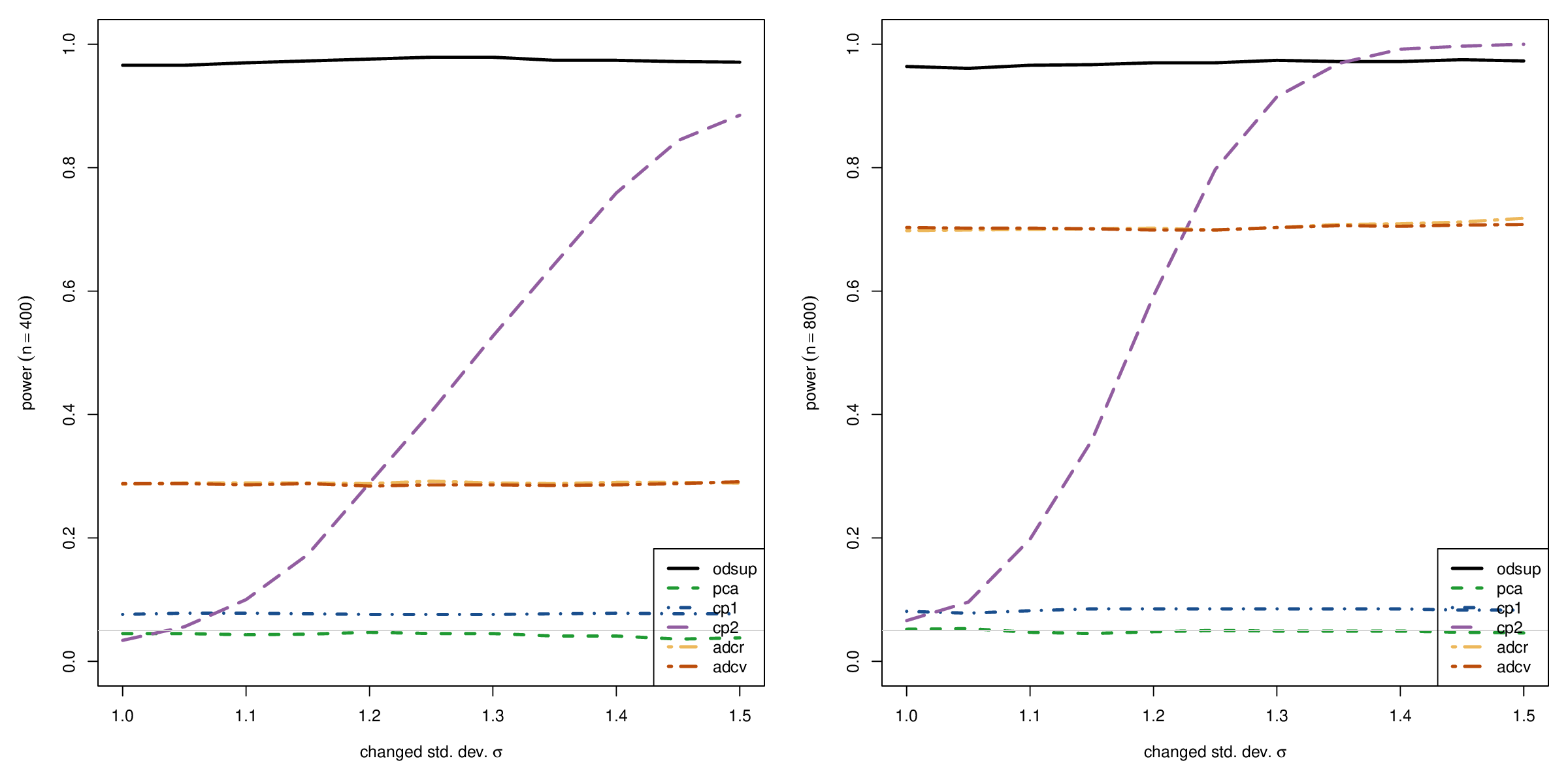}
\caption{Empirical power curves of six \textsc{iid} tests in different models (Row 1: MD and VCP; Row 2: AR and MA; Row 3: MDMA and VCPMA with moving average coefficient $b=7$; Row 4: MDMA and VCPMA with moving average coefficient $b=10$) with different sample sizes ($400$ and $800$) based on $1000$ Monte Carlo replications}
\label{fig:power}
\end{figure}

\subsection{Real data examples}
We analyze a dataset of air pollutants from UCI machine learning repository (\url{https://doi.org/10.24432/C5RK5G}), the MNIST dataset \citep{lecun1998gradient} obtained from the \textsf{R} package \textbf{dslabs} (\url{https://cran.r-project.org/package=dslabs}) and the financial and email network data studied by \citet{billio2022bayesian}. 
The air pollutants data contain hourly observations of 6 main air pollutants at multiple sites in Beijing over the time period from March 1, 2013 to February 28, 2017. We focus on the first 400 observations without NA values from the Aotizhongxin station. 
The MNIST dataset has a large collection of handwritten digits, normalized to 28-by-28 images, from which the first 800 images in the training dataset are picked out. 
Moreover, we calculate the first left singular vector of each image and use these vectors as a transformed sample, denoted by MNIST\_V, mimicking the {\iid} setting. 
The financial network data consists of 2-layer 61-by-61 binary directed networks sampled at 110 time points, representing Granger-causal links about the return and realized volatility among 61 European financial institutions. 
The email data consists of 2-layer 90-by-90 binary directed networks sampled at 79 time points, representing the EUcore sender–receiver communication flows among 90 researchers at a European research institution in two departments. 

We carry out \textsc{odsup}, \textsc{cp1}, \textsc{cp2} and \textsc{adcv} to test whether the five samples obey the {\iid} assumption or not, using the same settings in Section~\ref{sec:simu} with vectorization, except that the kernel function is replaced by $h(x,y) = 1/(\norm{x-y}^{4}+1)$ for the sake of computational feasibility. 
The $p$-values are given in Table~\ref{tab:pval}, highlighting a practical advantage of our method. 
Remarkably, MNIST\_V is justified as {\iid} by all tests. 
The proposed test \textsc{odsup} is capable of detecting violations of the {\iid} property in all other datasets. 
By contrast, \textsc{cp1} and \textsc{cp2} retain the {\iid} assumption for the MNIST data that are plausibly autocorrelated, and so does \textsc{adcv} for the email network data that might contain distributional changes. 
The air pollutants data have possibly stable second moments, and thus \textsc{cp2} does not reject the {\iid} hypothesis within them. 
With significant departure from the {\iid} assumption, the financial network data obtain rejections by all tests. 
These results underscore again the adaptability and efficacy of the proposed method as a universal diagnostic tool prior to statistical modeling. 

\begin{table}[!ht]
\renewcommand{\arraystretch}{0.8}
    \centering
\caption{Results of four \textsc{iid} tests on five datasets}
\label{tab:pval}
\begin{tabular}{ccrrrr}
\hline
Data & Sample size & \textsc{odsup} $p$-value & \textsc{cp1} $p$-value & \textsc{cp2} $p$-value & \textsc{adcv} $p$-value \\
Air & 400 & 0.000 & 0.001 & 0.196 & 0.00 \\
MNIST & 800 & 0.000 & 0.084 & 0.081 & 0.00 \\
MNIST\_V & 800 & 0.569 & 0.699 & 0.197 & 0.10 \\
Financial & 110 & 0.000 & 0.000 & 0.000 & 0.00 \\
Email & 79 & 0.000 & 0.000 & 0.000 & 0.07 \\
\hline
\end{tabular}
\end{table}


\appendix
\section{Further Explanations of Theoretical Derivation}
\label{appn:theory}

For better intuition of the H\'{a}jek projection in our context, we concretize the notations in \eqref{eq:uproj-proc} and \eqref{eq:u0proj-proc} by the following examples of clustering and change-point, indicating the capabilities of detecting violations of the {\iid} assumption.
\begin{itemize}[leftmargin=*]
    \item Under $H_1^{\mathrm{cl}}$, one has 
    \[ \mu_n(t) = u_n(t) \E\{h(Y_1,Y_2)\} + \sum_{k=1}^{L}u_{nkk}(t) [\E\{h(Y_1,Y_1)\} - \E\{h(Y_1,Y_2)\}] ,\]
    \[ \varPi_{k}\{U_n(t)\} = u_{nkk}(t) [\varPi_{k}\{h(Y_k,Y_k)\} - 2 \varPi_{k}\{h(Y_k,Y_{L+1})\}] + \sum_{i\in\mathcal{I}_k} \nu_{ni}(t) \varPi_{k}\{h(Y_k,Y_{L+1})\} .\]
    Since $u_{nkk}(t) \lesssim m^{2} n^{-2}$, the mean function is nearly the same as in the {\iid} case, while the fluctuation induced by $Y_k$ is scaled as $\sum_{i\in\mathcal{I}_k} \nu_{ni}(t)$. Note that by definition \eqref{eq:coef_proj-null}, 
    \[ \abs{\nu_{ni}(t) - m^{-1}\nu_{\lfloor n/m \rfloor k}(t)} \lesssim m n^{-2} ,\quad (k-1)m < i \le \min(km,n) .\]
    It follows that 
    \[ \mu_n^\diamond(t) = \underbrace{\sum_{k=1}^{L}u_{nkk}^\diamond(t)}_{\order{m^{2} n^{-1}}} [\E\{h(Y_1,Y_1)\} - \E\{h(Y_1,Y_2)\}] ,\]
    \[ \varPi_{k}\{U_n^\diamond(t)\} = \underbrace{u_{nkk}^\diamond(t)}_{\order{m^{2} n^{-2}}} [\varPi_{k}\{h(Y_k,Y_k)\} - 2 \varPi_{k}\{h(Y_k,Y_{L+1})\}] + \underbrace{\sum_{i\in\mathcal{I}_k} \nu_{ni}^\diamond(t)}_{\nu_{\lfloor n/m \rfloor k}^\diamond(t) + \order{m^{2} n^{-2}}} \varPi_{k}\{h(Y_k,Y_{L+1})\} .\]
    In particular, $\mu_n^\diamond \equiv 0$ under the {\iid} hypothesis $H_0$. 
    \item Under $H_1^{\mathrm{cp}}$, writing $(Y'^< , Y'^>)$ as an independent copy of $(Y^< , Y^>)$, one has 
    \[\begin{aligned}
        \mu_n(t) &= \frac{n\tau^2-\tau}{n-1} u_{n\tau}\Big(\frac{t}{\tau}\Big) \E\{h(Y^<,Y'^<)\} + \frac{n(1-\tau)^2-(1-\tau)}{n-1} u_{n(1-\tau)}\Big(\frac{t}{1-\tau}\Big) \E\{h(Y^>,Y'^>)\} \\ &\quad + \bigg\{ u_n(t) - \frac{n\tau^2-\tau}{n-1} u_{n\tau}\Big(\frac{t}{\tau}\Big) - \frac{n(1-\tau)^2-(1-\tau)}{n-1} u_{n(1-\tau)}\Big(\frac{t}{1-\tau}\Big) \bigg\} \E\{h(Y^<,Y^>)\} , \\
        \mu_n^\diamond(t) &= \frac{n\tau^2-\tau}{n-1} \bigg\{u_{n\tau}\Big(\frac{t}{\tau}\Big) - u_n(t)\bigg\} [\E\{h(Y^<,Y'^<)\}-\E\{h(Y^<,Y^>)\}] \\ &\quad + \frac{n(1-\tau)^2-(1-\tau)}{n-1} \bigg\{u_{n(1-\tau)}\Big(\frac{t}{1-\tau}\Big) - u_n(t)\bigg\} [\E\{h(Y^>,Y'^>)\}-\E\{h(Y^<,Y^>)\}] ,
    \end{aligned}\]
    where $u_{n\tau},u_{n(1-\tau)}$ are extended as $1$ on $(1,\infty)$. 
    The norm $\norm{\mu_n^\diamond}_\infty$ thus reflects the change-point in terms of $\E\{h(Y^<,Y'^<)\}-\E\{h(Y^<,Y^>)\}$ and $\E\{h(Y^>,Y'^>)\}-\E\{h(Y^<,Y^>)\}$. 
    Besides, 
    \[ \varPi_{k}\{U_n(t)\} = \nu_{nk}(t) h_1^\diamond(X_k) ,\]
    \[ \varPi_{k}\{U_n^\diamond(t)\} = \nu_{nk}^\diamond(t) h_1^\diamond(X_k) ,\]
    provided that $\varPi_{k}\{h(X_k,Y^<)\} = \varPi_{k}\{h(X_k,Y^>)\} = h_1^\diamond(X_k)$. This corresponds to the case where the distributional changes of $h(X_i,X_j)$ are duly expressed via the expected values, e.g., $h(x,y) = x+y$ and the distributions of $Y^<,Y^>$ belong to a location family. 
\end{itemize}

In the aforementioned cases, one can see that 
\[ \Cov[\varPi_{k}\{U_n(s)\},\varPi_{k}\{U_n(t)\}] = \nu_{\lfloor n/m \rfloor k}(s) \nu_{\lfloor n/m \rfloor k}(t) \sigma_k^2 + \order{m^{2} n^{-2}} \]
for some $\sigma_k > 0$ representing the uncertainty within the $k$th cluster. 
Provided that $\lim_{n\to\infty} L \max_{1\le k\le L} |\sigma_k - \sigma| = 0$, the convergence in \eqref{eq:cov} will follow Lemma~\ref{lem:cov} below. 
\begin{lemma}\label{lem:cov}
Recall $\nu_{ni}$ defined in \eqref{eq:coef_proj-null}. Then 
\[ n\sum_{i=1}^{n}\nu_{ni}(s)\nu_{ni}(t) = \Gamma(s,t) + \order{n^{-1}} ,\]
where $\Gamma$ is the symmetric function defined on $[0,1]^2$ such that for $0\le s\le t\le 1$, 
\[ \Gamma(s,t) = \begin{cases}
4s(4t-2t^2-st-3^{-1}s^2) ,& s+t\le 1; \\
4\{s(1-s+2t-t^2)-3^{-1}(1-t)^3\} ,& s+t>1 .
\end{cases} \]
\end{lemma}
Similar results hold for the empirically centered process, and in particular,
\[ n\sum_{i=1}^{n}\nu_{ni}^\diamond(s)\nu_{ni}^\diamond(t) = \Gamma(s,t) - 4 u_\infty(s)u_\infty(t) + \order{n^{-1}} .\]

\section{Concrete Results of Simulation}
\label{appn:simu}
Results from the simulation are reported in the following Tables~\ref{tab:power}--\ref{tab:pow_cont} about the empirical size and power of the tests with significance level $\alpha=5\%$. 

\begin{table}[!ht]
\renewcommand{\arraystretch}{0.65}
    \centering\scriptsize
\caption{Rejection proportions (\%) calculated for six \textsc{iid} tests in various scenarios based on $1000$ Monte Carlo replications}
\label{tab:power}
    \begin{tabular}{c|rrrrrr|rrrrrr|}
\hline
\multicolumn{1}{c}{$\mathcal{M}_0$} & \multicolumn{6}{c}{$n=400$} & \multicolumn{6}{c}{$n=800$} \\
 & \textsc{odsup} & \textsc{pca} & \textsc{cp1} & \textsc{cp2} & \textsc{adcr} & \textsc{adcv} & \textsc{odsup} & \textsc{pca} & \textsc{cp1} & \textsc{cp2} & \textsc{adcr} & \textsc{adcv} \\
 & 5.6 & 5.3 & 3.4 & 4.1 & 2.4 & 2.8 & 5.0 & 4.6 & 5.2 & 5.7 & 3.2 & 3.0  \\
\hline
\multicolumn{1}{c}{MD} & \multicolumn{6}{c}{$n=400$} & \multicolumn{6}{c}{$n=800$} \\
$\mu \times 10^{4}$ & \textsc{odsup} & \textsc{pca} & \textsc{cp1} & \textsc{cp2} & \textsc{adcr} & \textsc{adcv} & \textsc{odsup} & \textsc{pca} & \textsc{cp1} & \textsc{cp2} & \textsc{adcr} & \textsc{adcv} \\
1.5 & 5.1 & 5.1 & 6.2 & 4.0 & 2.7 & 2.8 & 5.3 & 4.8 & 25.0 & 5.9 & 3.1 & 3.0  \\
3 & 5.5 & 5.1 & 13.8 & 3.9 & 2.6 & 2.8 & 11.6 & 5.9 & 78.7 & 9.9 & 4.7 & 4.5  \\
4.5 & 6.1 & 4.6 & 26.3 & 5.1 & 2.6 & 2.5 & 36.0 & 5.9 & 99.4 & 19.2 & 13.6 & 13.5  \\
6 & 8.3 & 4.5 & 46.4 & 6.2 & 2.8 & 2.7 & 76.1 & 5.1 & 100.0 & 46.7 & 44.9 & 44.9  \\
7.5 & 13.7 & 4.6 & 67.5 & 8.1 & 3.6 & 3.5 & 97.3 & 5.2 & 100.0 & 82.5 & 88.3 & 88.2  \\
9 & 21.5 & 4.9 & 82.5 & 11.4 & 5.1 & 5.0 & 100.0 & 4.9 & 100.0 & 98.5 & 99.9 & 99.9  \\
10.5 & 34.7 & 4.8 & 92.4 & 17.6 & 8.1 & 8.0 & 100.0 & 5.1 & 100.0 & 100.0 & 100.0 & 100.0 \\
12 & 49.2 & 4.3 & 98.4 & 26.4 & 14.7 & 15.3 & 100.0 & 4.8 & 100.0 & 100.0 & 100.0 & 100.0 \\
13.5 & 67.1 & 4.0 & 99.8 & 39.8 & 27.1 & 27.1 & 100.0 & 4.6 & 100.0 & 100.0 & 100.0 & 100.0 \\
15 & 81.2 & 4.2 & 100.0 & 53.8 & 43.6 & 43.5 & 100.0 & 4.7 & 100.0 & 100.0 & 100.0 & 100.0 \\
\hline
\multicolumn{1}{c}{VCP} & \multicolumn{6}{c}{$n=400$} & \multicolumn{6}{c}{$n=800$} \\
$\sigma$ & \textsc{odsup} & \textsc{pca} & \textsc{cp1} & \textsc{cp2} & \textsc{adcr} & \textsc{adcv} & \textsc{odsup} & \textsc{pca} & \textsc{cp1} & \textsc{cp2} & \textsc{adcr} & \textsc{adcv} \\
1.05 & 5.7 & 5.3 & 3.7 & 15.8 & 2.7 & 2.8 & 4.9 & 4.1 & 5.9 & 23.2 & 3.5 & 3.3    \\
1.1 & 7.1 & 5.5 & 4.6 & 43.6 & 2.9 & 2.7 & 7.5 & 4.1 & 6.5 & 70.6 & 3.7 & 3.5   \\
1.15 & 10.7 & 5.4 & 5.1 & 77.2 & 3.1 & 3.0 & 14.4 & 3.9 & 7.1 & 96.9 & 4.0 & 3.8   \\
1.2 & 20.6 & 5.4 & 6.3 & 94.5 & 3.1 & 3.2 & 32.2 & 3.9 & 8.0 & 99.9 & 4.5 & 4.2   \\
1.25 & 38.1 & 4.9 & 7.1 & 99.1 & 3.4 & 3.4 & 62.6 & 3.6 & 8.4 & 100.0 & 5.6 & 5.6  \\
1.3 & 60.4 & 4.6 & 7.9 & 99.9 & 3.8 & 3.7 & 87.5 & 3.5 & 9.1 & 100.0 & 7.0 & 6.8   \\
1.35 & 79.8 & 4.8 & 9.0 & 100.0 & 4.1 & 4.2 & 98.8 & 3.2 & 10.2 & 100.0 & 10.7 & 10.8 \\
1.4 & 93.2 & 4.4 & 10.2 & 100.0 & 5.5 & 5.5 & 100.0 & 3.0 & 11.1 & 100.0 & 17.0 & 16.9 \\
1.45 & 98.0 & 4.6 & 11.1 & 100.0 & 7.5 & 7.7 & 100.0 & 2.7 & 12.3 & 100.0 & 28.8 & 28.6 \\
1.5 & 99.6 & 4.6 & 12.2 & 100.0 & 11.4 & 10.8 & 100.0 & 2.8 & 13.7 & 100.0 & 47.2 & 48.0 \\
\hline
\multicolumn{1}{c}{AR} & \multicolumn{6}{c}{$n=400$} & \multicolumn{6}{c}{$n=800$} \\
$a \times 10$ & \textsc{odsup} & \textsc{pca} & \textsc{cp1} & \textsc{cp2} & \textsc{adcr} & \textsc{adcv} & \textsc{odsup} & \textsc{pca} & \textsc{cp1} & \textsc{cp2} & \textsc{adcr} & \textsc{adcv} \\
1 & 5.3 & 5.5 & 3.2 & 4.1 & 51.0 & 50.9 & 5.5 & 4.8 & 5.0 & 6.1 & 94.9 & 95.4   \\
2 & 5.8 & 5.2 & 2.6 & 7.0 & 100.0 & 100.0 & 6.5 & 4.2 & 3.9 & 8.1 & 100.0 & 100.0 \\
3 & 7.2 & 4.7 & 1.8 & 9.3 & 100.0 & 100.0 & 8.5 & 3.8 & 2.7 & 10.3 & 100.0 & 100.0 \\
4 & 10.4 & 2.9 & 1.0 & 13.7 & 100.0 & 100.0 & 10.7 & 2.8 & 1.4 & 14.7 & 100.0 & 100.0 \\
5 & 14.5 & 1.6 & 0.5 & 21.6 & 100.0 & 100.0 & 15.0 & 1.3 & 0.8 & 22.4 & 100.0 & 100.0 \\
6 & 20.8 & 0.1 & 0.2 & 32.2 & 100.0 & 100.0 & 20.7 & 0.2 & 0.3 & 33.2 & 100.0 & 100.0 \\
7 & 29.7 & 0.0 & 0.2 & 48.4 & 100.0 & 100.0 & 27.4 & 0.0 & 0.0 & 50.1 & 100.0 & 100.0 \\
8 & 65.9 & 0.0 & 0.0 & 67.4 & 100.0 & 100.0 & 44.1 & 0.0 & 0.0 & 71.0 & 100.0 & 100.0 \\
9 & 100.0 & 0.0 & 0.0 & 87.5 & 100.0 & 100.0 & 100.0 & 0.0 & 0.0 & 90.4 & 100.0 & 100.0 \\
10 & 100.0 & 0.0 & 0.0 & 100.0 & 100.0 & 100.0 & 100.0 & 0.0 & 0.0 & 100.0 & 100.0 & 100.0 \\
\hline
\multicolumn{1}{c}{MA} & \multicolumn{6}{c}{$n=400$} & \multicolumn{6}{c}{$n=800$} \\
$b$ & \textsc{odsup} & \textsc{pca} & \textsc{cp1} & \textsc{cp2} & \textsc{adcr} & \textsc{adcv} & \textsc{odsup} & \textsc{pca} & \textsc{cp1} & \textsc{cp2} & \textsc{adcr} & \textsc{adcv} \\
1 & 18.9 & 0.1 & 28.2 & 12.4 & 100.0 & 100.0 & 15.3 & 0.2 & 28.8 & 16.8 & 100.0 & 100.0 \\
2 & 17.9 & 0.8 & 23.3 & 8.4 & 100.0 & 100.0 & 13.6 & 0.4 & 23.8 & 11.4 & 100.0 & 100.0 \\
3 & 24.3 & 2.6 & 17.2 & 6.5 & 100.0 & 100.0 & 14.9 & 2.5 & 18.8 & 9.0 & 100.0 & 100.0 \\
4 & 41.0 & 3.3 & 13.3 & 4.6 & 100.0 & 100.0 & 26.0 & 3.8 & 15.3 & 8.5 & 100.0 & 100.0 \\
5 & 63.0 & 3.9 & 11.2 & 4.3 & 98.5 & 98.8 & 43.7 & 4.5 & 12.2 & 7.8 & 100.0 & 100.0 \\
6 & 82.4 & 4.2 & 9.5 & 4.1 & 88.9 & 89.7 & 67.3 & 4.5 & 11.3 & 7.1 & 100.0 & 100.0 \\
7 & 90.4 & 3.9 & 9.0 & 4.0 & 67.9 & 69.2 & 82.4 & 4.9 & 10.5 & 6.8 & 99.1 & 99.1  \\
8 & 94.0 & 4.1 & 8.3 & 3.8 & 51.6 & 51.7 & 90.6 & 5.2 & 9.1 & 6.5 & 93.5 & 93.4  \\
9 & 96.0 & 4.3 & 7.9 & 3.5 & 37.3 & 37.3 & 95.0 & 5.3 & 8.6 & 6.7 & 82.7 & 82.8  \\
10 & 96.6 & 4.5 & 7.6 & 3.4 & 28.7 & 28.8 & 96.4 & 5.2 & 8.1 & 6.6 & 69.8 & 70.3 \\
\hline
    \end{tabular}
\end{table}
\begin{table}[!ht]
\renewcommand{\arraystretch}{0.6}
    \centering\scriptsize
\caption{Rejection proportions (\%) calculated for six \textsc{iid} tests in various scenarios based on $1000$ Monte Carlo replications (Continued)}
\label{tab:pow_cont}
    \begin{tabular}{c|rrrrrr|rrrrrr|}
\hline
\multicolumn{1}{c}{MDMA} & \multicolumn{6}{c}{$n=400$} & \multicolumn{6}{c}{$n=800$} \\
\multicolumn{1}{c}{$b=7$} \\
$\mu \times 10^{3}$ & \textsc{odsup} & \textsc{pca} & \textsc{cp1} & \textsc{cp2} & \textsc{adcr} & \textsc{adcv} & \textsc{odsup} & \textsc{pca} & \textsc{cp1} & \textsc{cp2} & \textsc{adcr} & \textsc{adcv} \\
0 & 90.4 & 3.9 & 9.0 & 4.0 & 67.9 & 69.2 & 82.4 & 4.9 & 10.5 & 6.8 & 99.1 & 99.1 \\
1.5 & 89.8 & 4.1 & 14.9 & 4.3 & 70.2 & 70.4 & 82.9 & 4.1 & 54.5 & 9.2 & 99.2 & 99.2   \\
3 & 89.1 & 4.1 & 34.2 & 5.4 & 72.0 & 72.5 & 86.6 & 3.7 & 98.9 & 19.8 & 99.9 & 99.9   \\
4.5 & 91.2 & 4.1 & 60.5 & 8.0 & 76.5 & 76.5 & 91.9 & 4.5 & 100.0 & 61.6 & 100.0 & 100.0 \\
6 & 91.3 & 3.9 & 81.2 & 13.1 & 81.4 & 81.9 & 96.8 & 4.4 & 100.0 & 95.8 & 100.0 & 100.0 \\
7.5 & 92.3 & 3.5 & 93.6 & 21.3 & 87.2 & 87.5 & 99.3 & 4.5 & 100.0 & 100.0 & 100.0 & 100.0 \\
9 & 92.6 & 3.0 & 99.1 & 36.4 & 92.5 & 92.4 & 100.0 & 4.4 & 100.0 & 100.0 & 100.0 & 100.0 \\
10.5 & 94.0 & 3.1 & 100.0 & 58.0 & 96.5 & 96.6 & 100.0 & 4.4 & 100.0 & 100.0 & 100.0 & 100.0 \\
12 & 94.9 & 3.2 & 100.0 & 77.1 & 99.3 & 99.3 & 100.0 & 4.3 & 100.0 & 100.0 & 100.0 & 100.0 \\
13.5 & 96.4 & 3.1 & 100.0 & 91.2 & 99.9 & 99.9 & 100.0 & 4.2 & 100.0 & 100.0 & 100.0 & 100.0 \\
15 & 97.2 & 2.9 & 100.0 & 96.9 & 100.0 & 100.0 & 100.0 & 4.2 & 100.0 & 100.0 & 100.0 & 100.0 \\
\hline
\multicolumn{1}{c}{MDMA} & \multicolumn{6}{c}{$n=400$} & \multicolumn{6}{c}{$n=800$} \\
\multicolumn{1}{c}{$b=10$} \\
$\mu \times 10^{3}$ & \textsc{odsup} & \textsc{pca} & \textsc{cp1} & \textsc{cp2} & \textsc{adcr} & \textsc{adcv} & \textsc{odsup} & \textsc{pca} & \textsc{cp1} & \textsc{cp2} & \textsc{adcr} & \textsc{adcv} \\
0 & 96.6 & 4.5 & 7.6 & 3.4 & 28.7 & 28.8 & 96.4 & 5.2 & 8.1 & 6.6 & 69.8 & 70.3   \\
1.5 & 96.4 & 4.8 & 10.1 & 3.8 & 29.2 & 29.3 & 95.6 & 4.4 & 31.4 & 7.5 & 71.8 & 72.2  \\
3 & 96.8 & 4.2 & 17.8 & 4.3 & 30.2 & 30.1 & 96.1 & 4.3 & 80.1 & 11.2 & 78.0 & 78.0  \\
4.5 & 97.8 & 4.1 & 34.2 & 5.3 & 31.1 & 31.9 & 96.7 & 4.2 & 99.4 & 21.8 & 88.9 & 88.7  \\
6 & 96.7 & 4.0 & 53.5 & 7.0 & 34.7 & 34.6 & 96.6 & 4.7 & 100.0 & 49.4 & 97.8 & 97.7  \\
7.5 & 97.1 & 4.2 & 71.4 & 9.4 & 38.1 & 37.9 & 97.7 & 4.7 & 100.0 & 83.3 & 99.9 & 99.9  \\
9 & 96.6 & 4.4 & 84.5 & 14.0 & 44.5 & 45.1 & 98.1 & 4.9 & 100.0 & 98.4 & 100.0 & 100.0 \\
10.5 & 96.5 & 4.0 & 92.9 & 19.4 & 52.9 & 53.2 & 98.5 & 5.0 & 100.0 & 100.0 & 100.0 & 100.0 \\
12 & 97.0 & 3.6 & 98.0 & 29.3 & 63.2 & 64.2 & 99.5 & 4.4 & 100.0 & 100.0 & 100.0 & 100.0 \\
13.5 & 96.9 & 3.5 & 99.7 & 41.5 & 73.4 & 73.9 & 99.6 & 4.5 & 100.0 & 100.0 & 100.0 & 100.0 \\
15 & 97.2 & 3.6 & 100.0 & 57.5 & 84.3 & 84.5 & 99.9 & 4.5 & 100.0 & 100.0 & 100.0 & 100.0 \\
\hline
\multicolumn{1}{c}{VCPMA} & \multicolumn{6}{c}{$n=400$} & \multicolumn{6}{c}{$n=800$} \\
\multicolumn{1}{c}{$b=7$} \\
$\sigma$ & \textsc{odsup} & \textsc{pca} & \textsc{cp1} & \textsc{cp2} & \textsc{adcr} & \textsc{adcv} & \textsc{odsup} & \textsc{pca} & \textsc{cp1} & \textsc{cp2} & \textsc{adcr} & \textsc{adcv} \\
1 & 90.4 & 3.9 & 9.0 & 4.0 & 67.9 & 69.2 & 82.4 & 4.9 & 10.5 & 6.8 & 99.1 & 99.1 \\
1.05 & 92.0 & 4.3 & 8.9 & 6.2 & 68.1 & 69.7 & 85.6 & 5.1 & 10.6 & 10.1 & 98.9 & 99.1 \\
1.1 & 93.5 & 4.3 & 9.1 & 10.4 & 68.5 & 69.7 & 88.2 & 4.6 & 10.5 & 20.7 & 98.9 & 99.0 \\
1.15 & 94.7 & 4.7 & 9.2 & 17.6 & 68.8 & 69.4 & 91.0 & 4.3 & 10.6 & 36.0 & 99.0 & 99.0 \\
1.2 & 94.9 & 4.5 & 9.1 & 29.2 & 69.4 & 69.8 & 92.4 & 4.5 & 10.8 & 59.4 & 98.9 & 98.9 \\
1.25 & 95.4 & 4.4 & 9.1 & 40.7 & 69.0 & 69.8 & 93.3 & 4.4 & 10.7 & 79.6 & 98.9 & 99.0 \\
1.3 & 95.9 & 4.3 & 9.3 & 52.6 & 69.0 & 69.9 & 94.0 & 4.5 & 10.9 & 91.4 & 98.8 & 99.0 \\
1.35 & 96.0 & 3.9 & 9.3 & 64.5 & 69.0 & 69.8 & 94.9 & 4.6 & 11.0 & 96.5 & 98.7 & 98.9 \\
1.4 & 96.5 & 4.0 & 9.3 & 75.3 & 68.9 & 69.5 & 95.8 & 4.9 & 10.8 & 99.2 & 98.9 & 99.0 \\
1.45 & 96.8 & 4.3 & 9.3 & 83.7 & 69.2 & 69.4 & 96.2 & 4.8 & 10.8 & 99.7 & 98.9 & 99.1 \\
1.5 & 97.1 & 3.9 & 9.2 & 88.4 & 69.4 & 69.6 & 96.3 & 4.8 & 10.9 & 100.0 & 98.9 & 99.1 \\
\hline
\multicolumn{1}{c}{VCPMA} & \multicolumn{6}{c}{$n=400$} & \multicolumn{6}{c}{$n=800$} \\
\multicolumn{1}{c}{$b=10$} \\
$\sigma$ & \textsc{odsup} & \textsc{pca} & \textsc{cp1} & \textsc{cp2} & \textsc{adcr} & \textsc{adcv} & \textsc{odsup} & \textsc{pca} & \textsc{cp1} & \textsc{cp2} & \textsc{adcr} & \textsc{adcv} \\
1 & 96.6 & 4.5 & 7.6 & 3.4 & 28.7 & 28.8 & 96.4 & 5.2 & 8.1 & 6.6 & 69.8 & 70.3 \\
1.05 & 96.6 & 4.5 & 7.8 & 5.6 & 28.9 & 28.8 & 96.1 & 5.3 & 7.8 & 9.6 & 69.9 & 70.2 \\
1.1 & 97.0 & 4.3 & 7.8 & 10.0 & 28.9 & 28.6 & 96.6 & 4.7 & 8.2 & 19.8 & 70.0 & 70.2 \\
1.15 & 97.3 & 4.4 & 7.7 & 17.3 & 28.9 & 28.8 & 96.7 & 4.5 & 8.5 & 35.8 & 70.1 & 70.1 \\
1.2 & 97.6 & 4.7 & 7.6 & 28.9 & 28.8 & 28.4 & 97.0 & 4.8 & 8.5 & 59.2 & 70.2 & 69.9 \\
1.25 & 97.9 & 4.5 & 7.6 & 40.4 & 29.2 & 28.6 & 97.0 & 5.0 & 8.5 & 79.7 & 69.9 & 69.9 \\
1.3 & 97.9 & 4.5 & 7.6 & 52.7 & 28.9 & 28.6 & 97.4 & 4.9 & 8.5 & 91.5 & 70.3 & 70.3 \\
1.35 & 97.4 & 4.1 & 7.7 & 64.4 & 28.8 & 28.5 & 97.2 & 4.9 & 8.5 & 96.9 & 70.8 & 70.6 \\
1.4 & 97.4 & 4.1 & 7.8 & 75.9 & 29.0 & 28.6 & 97.2 & 4.9 & 8.5 & 99.2 & 70.9 & 70.5 \\
1.45 & 97.2 & 3.6 & 7.7 & 84.4 & 29.0 & 28.8 & 97.5 & 4.7 & 8.3 & 99.7 & 71.2 & 70.7 \\
1.5 & 97.1 & 3.8 & 7.7 & 88.5 & 28.9 & 29.1 & 97.3 & 4.6 & 8.3 & 100.0 & 71.8 & 70.8 \\
\hline
    \end{tabular}
\end{table}

\bigskip
\begin{center}
{\large\bf SUPPLEMENTARY MATERIAL}
\end{center}

\begin{description}
\item[supp] Auxiliary theoretical results and proofs for all theorems, corollaries and lemmas in this paper. 
\item[code] 
The R-code for producing results in Section~\ref{sec:numeric} is publicly available at the GitHub repository \url{https://github.com/kellty/TestIID}. 
\end{description}


\bibliographystyle{agsm}
\bibliography{ref}

\end{document}